\documentclass[12pt]{article}
\usepackage{epsfig,amsmath,amssymb,euscript,psfrag,cite,color}

\def\pslash{\rlap{\hspace{0.02cm}/}{p}}

\def\sslash{\rlap{\hspace{0.02cm}/}{s}}

\def\tr{\hbox{tr}}
\def\prob{\hbox{Prob}}

\def\mytauone{(p_\tau \cdot p_1)}
\def\mytautwo{(p_\tau \cdot p_2)}
\def\mytauthree{(p_\tau \cdot p_3)}
\def\myonetwo{(p_1 \cdot p_2)}

\def\mysone{(s \cdot p_1)}
\def\mystwo{(s \cdot p_2)}
\def\mysthree{(s \cdot p_3)}
\def\myeps{(\varepsilon_{\alpha\beta\gamma\delta} p_2^\alpha p_3^\beta p_\tau^\gamma s^\delta)}

\def\ee{\epsilon}

\def\bm#1{\mbox{\boldmath$#1$\unboldmath}}
\def\comma{\quad , \quad}

\begin{document}

\begin{titlepage}

\begin{flushright}
SI-HEP-2015-07 \\
QFET-2015-08 \\
MITP/15-022 \\[3mm]

June 25, 2015
\end{flushright}
\begin{center}
\Large\bf 
Angular Analysis of New Physics Operators \\
  in polarized $\tau \to 3 \ell$ Decays

\end{center}
\begin{center}
{\sc Robin Br\"user$^a$,
Thorsten Feldmann$^a$,  Bj\"orn O.~Lange$^a$, \\
Thomas Mannel$^a$,  Sascha Turczyk$^b$}\\
\vspace{0.7cm}
{\sl $^a$ Theoretische Elementarteilchenphysik,
  Naturwissenschaftlich-technische Fakult\"at,
Universit\"at Siegen, 57068 Siegen, Germany } \\[0.25em]
{\sl $^b$ PRISMA Cluster of Excellence \& Mainz Institute for
  Theoretical Physics, \\ Johannes Gutenberg University, 55099 Mainz, Germany } \\
\end{center}

\begin{abstract}
  \vspace{0.2cm}\noindent In a bottom-up approach we investigate
  lepton-flavour violating processes $\tau \to 3 \ell$ that are
  mediated by New Physics encoded in effective-theory operators of dimension
  six. While the opportunity to scrutinize the underlying operator
  structure has been investigated before, we explore the benefits of
  utilising the polarization direction of the initial $\tau$ lepton
  and the angular distribution of the decay. Given the rarity of these
  events (if observed at all), we focus on integrated observables
  rather than spectra, such as partial rates and asymmetries. In an
  effort to estimate the number of events required to extract the
  coupling coefficients to the effective operators we perform a
  phenomenological study with virtual experiments.

\end{abstract}
\vfil

\end{titlepage}

\section{Introduction}

Within the Standard Model (SM) of particle physics, lepton flavour is
conserved as long as the neutrino masses are exactly zero. The
discovery of massive neutrinos and neutrino oscillations, however,
shows that lepton flavour violation (LFV) is in principle allowed. For
example the process $\tau^- \to \mu^- \gamma$ can occur at the
one-loop level if a tau-neutrino oscillates into a muon-neutrino
within the loop that is supplemented with a charged vector boson. If
this mechanism were the only source of LFV then the branching
fractions of the $\tau^- \to \mu^- \ell^+ \ell^-$ and $\tau^- \to e^-
\ell^+ \ell^-$ decay channels are non-zero, but tiny -- $\mathcal
O(10^{-45\pm 5})$ or thereabouts -- and clearly unobservable. Many New
Physics (NP) scenarios, however, predict much higher branching
fractions, very roughly $\mathcal O(10^{-10})$, which are at the edge
of observability (see e.g.~the review in \cite{Raidal:2008jk}). Here
and below $\ell$ denotes either a muon or an electron.

If such a decay were to be discovered (see e.g.~\cite{Aaij:2014azz}),
as is not unrealistic given the hints on deviations from the SM in the
lepton sector with the recent measurements%
\footnote{The LHCb measurement $R_K = 0.745^{+0.097}_{-0.082}$ is
  $2.6\sigma$ away from the SM expectation $R_K = 1+\mathcal
  O(10^{-4})$ and challenges the notion of lepton-flavour
  universality.}  of $R_K = \mathcal{B}(B\to K \mu^+ \mu^-) /
\mathcal{B}(B\to K e^+ e^-)$ by the LHCb collaboration
\cite{Aaij:2014ora} and the LFV Higgs decay $\mathcal{B}(h \to \mu
\tau) = (0.89 ^{+0.40}_{-0.37}) \%$ by the CMS experiment
\cite{CMS:2014hha}, it will be highly interesting to investigate the
underlying interaction structure in order to disentangle possible NP
models. In this paper we employ a bottom-up approach and treat the SM
as an effective field theory, i.e.~we consider higher-dimensional
operators that consist only of SM fields and respect SM symmetries.
Our aim is to gain qualitative and quantitative information on the
couplings of these operators. It is obvious that such an approach will
allow us to only gain direct insight into the couplings at the {\it
  low scale} -- set by the tau-lepton mass -- and not at the high
scale of the responsible NP mechanisms.  The implementation of the
renormalization-group running of the couplings, which has been derived
at one-loop level in
\cite{Jenkins:2013zja,Jenkins:2013wua,Alonso:2013hga}, is beyond the
scope of the current paper.

As some of us have shown in a previous publication
\cite{Dassinger:2007ru}, the $\tau \to 3\ell$ decays are mediated by a
handful of dimension-6 operators with possibly complex
coefficients. In that publication a Dalitz-plot analysis was
entertained in order to differentiate between radiative and leptonic
operators. Clearly, a reconstruction of a Dalitz distribution requires
a large data sample which obviously is hard to obtain for a very rare,
lepton number violating decay.  Thus, in order to obtain information
on the type of operator that mediates the decay and on the helicity
structure of the interaction, appropriate observables need to be
defined.  Obvious candidates are observables such as forward-backward
asymmetries, which can be measured even at very low statistics. Our
strategy is therefore to define observables of partially integrated
phase space, which can be measured by a simple counting experiment.

  Additional information on the structure of the interaction can be
  gained by studying the decay of polarized tau leptons
  \cite{Mannel:2014sba}. Such a polarization can be realized at $e^+
  e^-$ colliders \cite{Tsai:1994rc} running close to the $\tau^+
  \tau^-$ threshold with polarized electrons or electrons and
  positrons. As we shall show in this paper, taking into account the
  spin direction of the decaying tau lepton allows us to obtain
  information on the structure of the interaction, even with quite
  small data samples.

  The idea of using polarized tau leptons has also been discussed in
  \cite{Goto:2010sn}, however, in the context of a specific NP
  model. Our focus is a general analysis in order to pin down the
  structure of the relevant interaction.

Considering the possible LFV tau decays into leptons one may classify
the six distinct channels as (a) $\tau^- \to \ell^- \ell^- \ell^+$ with
$\ell = \mu, e$, or (b) $ \tau^- \to \mu^- e^- e^+, \tau^- \to e^- \mu^-
\mu^+$, or (c) $\tau^- \to e^- e^- \mu^+, \tau^- \to \mu^- \mu^- e^+$. The
processes in $(a)$ and $(c)$ involve identical particles in the final
state, whereas $(b)$ does not. Radiative operators contribute to
classes $(a)$ and $(b)$, but not $(c)$. For definiteness and simplicity
we focus our attention solely on case $(a)$, in which two external mass scales
suffice, the mass of the tau lepton, $m_\tau$, and the mass of the lighter
lepton, $m$. 

The paper is organised as follows: In the next section we define the
effective Hamiltonian with open coefficients $\xi_i$, where $i$ counts
through the various operators of mass-dimension 6. The setup of our
calculation is described in detail, including the polarisation vector
of the initial tau and the angles characterizing the position of the
polarisation vector relative to the decay plane. As a result the
totally differential decay rate is decomposed in terms of
trigonometric functions of these angles.  In chapter~\ref{sec:3} we
integrate the differential decay rate over the relevant parts of the
phase space. The results are rather bulky in print, so we have
diverted them to the appendix for easier reading. Chapter~\ref{sec:4}
consists of a phenomenological study of the two decay channels
$\tau \to 3\mu$ and $\tau \to 3e$. The main question we try to answer
is how well one could determine the Wilson coefficients of the
dimension-6 operators under the hypothesis of having discovered a low
number of events experimentally. We summarize in chapter~\ref{sec:5}.

\section{Calculational framework}
\subsection{Operator basis}

In the following, we adopt the conventions and notation of
\cite{Dassinger:2007ru}. (For more details we also refer the reader to that
reference.). Starting point is the most general set of 
dimension-six operators respecting the SM gauge symmetries
(see \cite{Buchmuller:1985jz,Grzadkowski:2010es}). After
integrating out the weak gauge bosons and the Higgs field after electroweak
symmetry breaking, there remain
four purely leptonic 4-fermion operators of dimension six.
Ordering the operators by the chirality of the involved lepton fields,
they can be expressed as
\begin{eqnarray}
  H^{(LL)(LL)}_{\rm lept} &=& \frac{g^{(LL)(LL)}_V}{\Lambda^2} (\bar \ell_L
  \gamma_\mu \tau_L)(\bar \ell_L \gamma^\mu \ell_L) \comma
  H^{(RR)(RR)}_{\rm lept} = \frac{g^{(RR)(RR)}_V}{\Lambda^2} (\bar \ell_R
  \gamma_\mu \tau_R)(\bar \ell_R \gamma^\mu \ell_R) \;, \nonumber \\
  H^{(LL)(RR)}_{\rm lept} &=& \frac{g^{(LL)(RR)}_V}{\Lambda^2} (\bar \ell_L
  \gamma_\mu \tau_L)(\bar \ell_R \gamma^\mu \ell_R) \comma
  H^{(RR)(LL)}_{\rm lept} = \frac{g^{(RR)(LL)}_V}{\Lambda^2} (\bar \ell_R
  \gamma_\mu \tau_R)(\bar \ell_L \gamma^\mu \ell_L) \;,
  \cr &&
  \label{eq:1}
\end{eqnarray}
where we have already singled out the $\tau$-lepton fields.  In this
notation, the chirality structure $(LR)(RL)$ can be Fierz transformed
to contribute to $H^{(LL)(RR)}_{\rm lept}$ and $H^{(RR)(LL)}_{\rm
  lept}$.  Notice that in order to feed terms of the form $(LR)(LR)$,
one would have to include dimension-eight operators in the SM
effective field theory.  Following \cite{Dassinger:2007ru}, we assume
that the NP scale $\Lambda$ is sufficiently large compared to the
electroweak scale, that these contributions can be neglected (for
similar reasoning in other context, see e.g.\
\cite{Alonso:2014csa,Cata:2015lta}).  In an explicit UV completion of
the SM, the dimensionless couplings $g_V$ 
should be determined from a matching calculation. \\

Similarly, one generates radiative operators, where, 
at low energies, only couplings to the photon field have to be considered.
(Contributions from intermediate $W^\pm$ or $Z^0$ bosons are already contained 
in (\ref{eq:1}).)
We are then left with two more terms in
the effective Hamiltonian, 
\begin{equation}
H_{\rm rad}^{(LR)} = - \frac{e}{8\pi} \frac{v}{\Lambda^2} g_{\rm
    rad}^{(LR)} (\bar \ell_L \sigma_{\mu\nu} \tau_R ) F^{\mu \nu}
  \comma
H_{\rm rad}^{(RL)} = - \frac{e}{8\pi} \frac{v}{\Lambda^2} g_{\rm
    rad}^{(RL)} (\bar \ell_R \sigma_{\mu\nu} \tau_L ) F^{\mu \nu}
\;.\nonumber
\end{equation}
They contribute to the $\tau^- \to \ell^- \ell^- \ell^+$ amplitude via photon exchange,
schematically 
\begin{eqnarray}
  \langle \ell \ell \bar\ell | H_{\rm rad}^{(LR)} | \tau \rangle 
  &=& \underbrace{ \alpha_{\rm em} \frac{v}{\Lambda^2} g_{\rm
    rad}^{(LR)}}_{\xi_5 m_\tau} \left( \bar\ell_L  \frac{(-i)}{q^2} q^\eta \sigma_{\mu
      \eta} \tau_R \right) \left( \bar \ell \gamma^\mu \ell \right)
  \;,   \nonumber \\
  \langle \ell \ell \bar\ell | H_{\rm rad}^{(RL)}| \tau  \rangle 
  &=& \underbrace{ \alpha_{\rm em} \frac{v}{\Lambda^2} g_{\rm
    rad}^{(RL)}}_{\xi_6 m_\tau} \left( \bar\ell_R  \frac{(-i)}{q^2} q^\eta \sigma_{\mu
      \eta} \tau_L \right) \left( \bar \ell \gamma^\mu \ell \right) \;.
  \label{eq:2}
\end{eqnarray}
Here, for simplicity, we used the same notation for lepton fields and on-shell spinors,
$q^\mu$ denotes the momentum flow through the
virtual photon, and $\alpha_{\rm em}$ the usual electromagnetic
fine-structure constant. 
In summary, the generic effective Hamiltonian for $\tau
\to 3 \ell$ decays can be written as
\begin{equation}
  \label{eq:3}
  H_{\rm eff} = H^{(LL)(LL)}_{\rm lept}  + H^{(RR)(RR)}_{\rm lept}  +
  H^{(LL)(RR)}_{\rm lept}  + H^{(RR)(LL)}_{\rm lept}  + 
  H_{\rm rad}^{(LR)} + H_{\rm rad}^{(RL)} \;.
\end{equation}
For future convenience we shall combine the six couplings into a
complex-valued vector 
\begin{equation}
  \label{eq:4}
  \xi^T = \left( \frac{g^{(LL)(LL)}_V}{\Lambda^2},
    \frac{g^{(RR)(RR)}_V}{\Lambda^2} ,
    \frac{g^{(LL)(RR)}_V}{\Lambda^2},
    \frac{g^{(RR)(LL)}_V}{\Lambda^2}, 
    \frac{\alpha_{\rm em}}{m_\tau} \frac{v}{\Lambda^2} g_{\rm
      rad}^{(LR)} , 
    \frac{\alpha_{\rm em}}{m_\tau} \frac{v}{\Lambda^2} g_{\rm
      rad}^{(RL)} \right)
\end{equation}
of mass dimension (-2).

\subsection{Spin polarization of the tau lepton}

The spin polarization of a beam of particles is typically measured in
the flight direction of said particles. In the setup of our
calculation we will define the $z$-axis of our lab-frame coordinate
system to be the flight direction of the tau lepton. 
The reference vector for the spin orientation is then chosen as 
$s^\mu=(0,0,0,1)$ in the tau lepton's rest frame, with 
$s^2 = -1$ and $s\cdot p_\tau = 0$. 
In the calculation of the squared amplitude $|\mathcal M|^2$
we will then use the Dirac matrix 
\begin{equation}
  \label{eq:6}
  u^\uparrow \bar u^\uparrow = (\pslash_\tau + m_\tau)
  \, \frac{1+\gamma_5 \, \sslash}{2} 
\end{equation}
to project onto tau leptons with ``spin-up''.
In this way the spin vector $s^\mu$ 
only appears at most linearly in the squared amplitude.
Note that there are then only three linearly independent 
invariants that one can build from $s^\mu$ and the lepton momenta
in the decay $\tau^-(p_\tau) \to
\ell^-(p_1) \, \ell^-(p_2) \,\ell^+(p_3)$. These can be taken as
\begin{equation}
  \label{eq:7}
    t = s\cdot p_2 \comma 
  u = s\cdot p_3 \comma
  v = \varepsilon_{\alpha\beta\gamma\delta} \ p_\tau^\alpha \, s^\beta
  \, p_2^\gamma \, p_3^\delta \;,
\end{equation}
and therefore the squared amplitude of the spin-up tau decay can be
decomposed as
\begin{eqnarray}
  \label{eq:8}
  |\mathcal M^\uparrow(p_\tau,p_1,p_2,p_3,s)|^2 &=& g_1(p_\tau,p_1,p_2,p_3) +
  g_t(p_\tau,p_1,p_2,p_3) \, t \nonumber \\
&& + g_u(p_\tau,p_1,p_2,p_3) \, u + g_v(p_\tau,p_1,p_2,p_3) \, v\;.
\end{eqnarray}
Similarly the decay of a ``spin-down'' polarized tau lepton can be
calculated with the help of the projector $u^\downarrow \bar
u^\downarrow = (\pslash_\tau + m_\tau) (1-\gamma_5 \sslash)/2$. We
stress that the direction in which the spin is measured, i.e.~the 
$z$-axis (or reference vector $s^\mu$) remains fixed. With this point of
view%
\footnote{Alternatively one may perform the transformation $s^\mu \to
  -s^\mu$ to find the differential decay rate of a ``spin-down''
  polarized tau.} 
the variables defined in (\ref{eq:7}) remain
unchanged when describing the decay of a spin-down rather than spin-up
tau lepton. In this spin-down case one finds the same functions
$g_{1,t,u,v}$ from above that describe the squared amplitude, albeit
in the combination
\begin{eqnarray}
  \label{eq:9}
    |\mathcal M^\downarrow(p_\tau,p_1,p_2,p_3,s)|^2 &=& g_1(p_\tau,p_1,p_2,p_3) -
  g_t(p_\tau,p_1,p_2,p_3) t \nonumber \\
&& - g_u(p_\tau,p_1,p_2,p_3) u - g_v(p_\tau,p_1,p_2,p_3) v\;.
\end{eqnarray}
Therefore only the part $2 g_1$ contributes to the {\it unpolarized}
decay rate. The information contained in the functions $g_{t,u,v}$
would then be lost and could not be used in the effort to unveil the
underlying operator structure. In what follows, 
we will use a slightly modified version of the
decomposition (\ref{eq:8}) based on angles associated with these invariants.

\subsection{Euler rotations}
The three momenta $\vec p_1, \vec p_2, \vec p_3$ of the decay product
span the decay plane. In general the spin direction $\vec s$ does not
lie in this decay plane, and its orientation can be described with the
help of two angles. For example, the angle between the normal of the
decay plane and $\vec s$ is the first, and the angle between the
momentum $p_3$ of the antilepton and the projection of $\vec s$ into
the decay plane the second angle. For convenience we employ the
technique of Euler rotations instead, which also serve to define the
orientation of $\vec s$ and the decay plane: \\

We define the lab frame as the reference frame (RF) in which the tau lepton
is at rest and the z-axis is aligned with the spin vector $\vec s =
\vec e_z$. 
We will now perform rotations until this plane is spanned by the new
basis vectors $\vec{e_x}'$ and $\vec{e_z}'$. Furthermore the
antilepton's momentum shall point directly in the $\vec{e_z}'$
direction. This new RF is called the decay frame, or RF'. We define
the following order of rotations to get from RF to RF':
\begin{equation}
  \label{eq:10}
  \vec{p_1}' = R_z(\alpha) R_y(\beta) R_z(\gamma) \, \vec p_1 \;, \quad
  \hbox{et cetera,}
\end{equation}
with mathematically positive convention:
\begin{equation}
  \label{eq:11}
  R_z(\alpha) = \left( \begin{array}{ccc}
\cos \alpha & - \sin \alpha & 0 \\
\sin \alpha & \cos \alpha & 0 \\
0 & 0 & 1
\end{array} \right) \comma
R_y(\beta) = \left( \begin{array}{ccc}
\cos \beta & 0 &  \sin \beta \\
0 & 1 & 0 \\
- \sin \beta & 0 & \cos \beta
\end{array} \right) \;.
\end{equation}
Hence the polarization direction vector in RF' reads
\begin{equation}
  \label{eq:12}
  \vec{s}^{\;\prime} = (\cos\alpha\, \sin\beta, \sin\alpha\,\sin\beta,
  \cos\beta) \;.
\end{equation}
Note that there is no dependence on the first rotation angle $\gamma$,
which reflects the azimuthal symmetry of the problem. 
The rotations are such that the decay plane coincides with the $x'-z'$
plane, so that we may parameterize
\begin{equation}
  \label{eq:13}
  \vec{p_1}' = - (\vec{p_2}' + \vec{p_3}') \comma
  \vec{p_2}' = \left( \begin{array}{c} b_2 \, \sin \varphi \\ 0 \\  b_2 \, \cos\varphi \end{array} \right) \comma  
  \vec{p_3}' = \left( \begin{array}{c} 0 \\ 0 \\ b_3 \end{array} \right) \;,
\end{equation}
where the parameter $\varphi = \phi_{23}$ has a physical interpretation of the angle
between $\vec p_2$ and $\vec p_3$ within the decay plane, which is
independent of the rotation angles $\alpha, \beta, \gamma$. The
lengths of the 3-vectors $b_2 = |\vec{p_2}'| = |\vec p_2|$ and
similarly $b_3$ are expressed in terms of the energies $E_i$ as
\begin{equation} 
b_2 = \sqrt{E_2^2 - m^2} \comma  
  b_3 = \sqrt{E_3^2 - m^2} \comma
\cos\phi_{23} = \frac{E_1^2- E_2^2 - E_3^2 + m^2}{2 \sqrt{E_2^2  - m^2}
  \sqrt{E_3^2 - m^2}}\;.  \label{eq:14}
\end{equation}
While the momentum of the antilepton, $p_3$, is well defined, the lepton
pair in the final state is indistinguishable if the leptons are of the same
flavour, and we must be careful not to double count. \\

For the last Euler rotation, it is sufficient to consider values of $\alpha \in [0,\pi)$ 
to lock the decay plane into the $x'-z'$ plane.
Therefore, in total we restrict the values of the rotation angles as
\begin{equation}
  \label{eq:15}
  \alpha \in [0,\pi) \comma \beta \in [0,\pi) \comma \gamma \in [0,2\pi)\;.
\end{equation}
Furthermore it is obvious that either $\vec{p_1}'$ has a positive $x'$
component and $\vec{p_2}'$ a negative one, or the other way
around. The parameter $\phi_{23}$ is well defined as the angle between
$\vec p_2$ and $\vec p_3$ if we insist on naming that lepton's
momentum $\vec p_2$ which has a positive%
\footnote{With this convention the invariant $v =
  \varepsilon_{\alpha\beta\gamma\delta} p_2^\alpha p_3^\beta
  p_\tau^\gamma s^\delta = \varepsilon_{\alpha\beta\gamma\delta}
  p_1^\alpha p_2^\beta p_3^\gamma s^\delta$ is negative definite.}
$x'$ component, i.e.
\begin{equation}
  \label{eq:16}
  \phi_{23} \in [0,\pi) \;.
\end{equation}
(However, the squared amplitudes, which we are going to calculate, will
be symmetric under the exchange of $p_1 \leftrightarrow p_2$, as they
must for identical particles in the final state.) \\
In terms of the angles and energies the invariants in (\ref{eq:7}) are
\begin{eqnarray}
  \label{eq:17}
  t &=& - b_2 \sin\phi_{23} \cos\alpha \sin \beta - b_2 \cos\phi_{23}
  \cos\beta \;, \nonumber \\
  u &=& - b_3 \cos\beta \;, \nonumber \\
  v &=& - m_\tau b_2 b_3 \sin\phi_{23} \sin\alpha \sin\beta\;.
\end{eqnarray}
Therefore we may decompose the squared amplitude in (\ref{eq:8}) in
terms of the trigonometric functions that appear above instead of the
invariants themselves. Specifically we have
\\[3mm]
\framebox{\begin{minipage}{0.99\textwidth}
\begin{equation}
  \label{eq:18}
  |\mathcal M^\uparrow|^2 = J_1(E_2,E_3) + J_2(E_2,E_3) \cos\beta + J_3(E_2,E_3)
  \cos \alpha \sin\beta + J_4(E_2,E_3) \sin \alpha  \sin\beta \;.
\end{equation}
$\left. \right.$
\end{minipage}} \\[3mm]

This setup allows us to quite easily pick out the individual
contributions from $J_1$ to $J_4$ by folding the differential
decay rate with appropriate weight functions.

\subsection{Phase space}
\begin{figure}[t]
\begin{center}
\epsfig{file=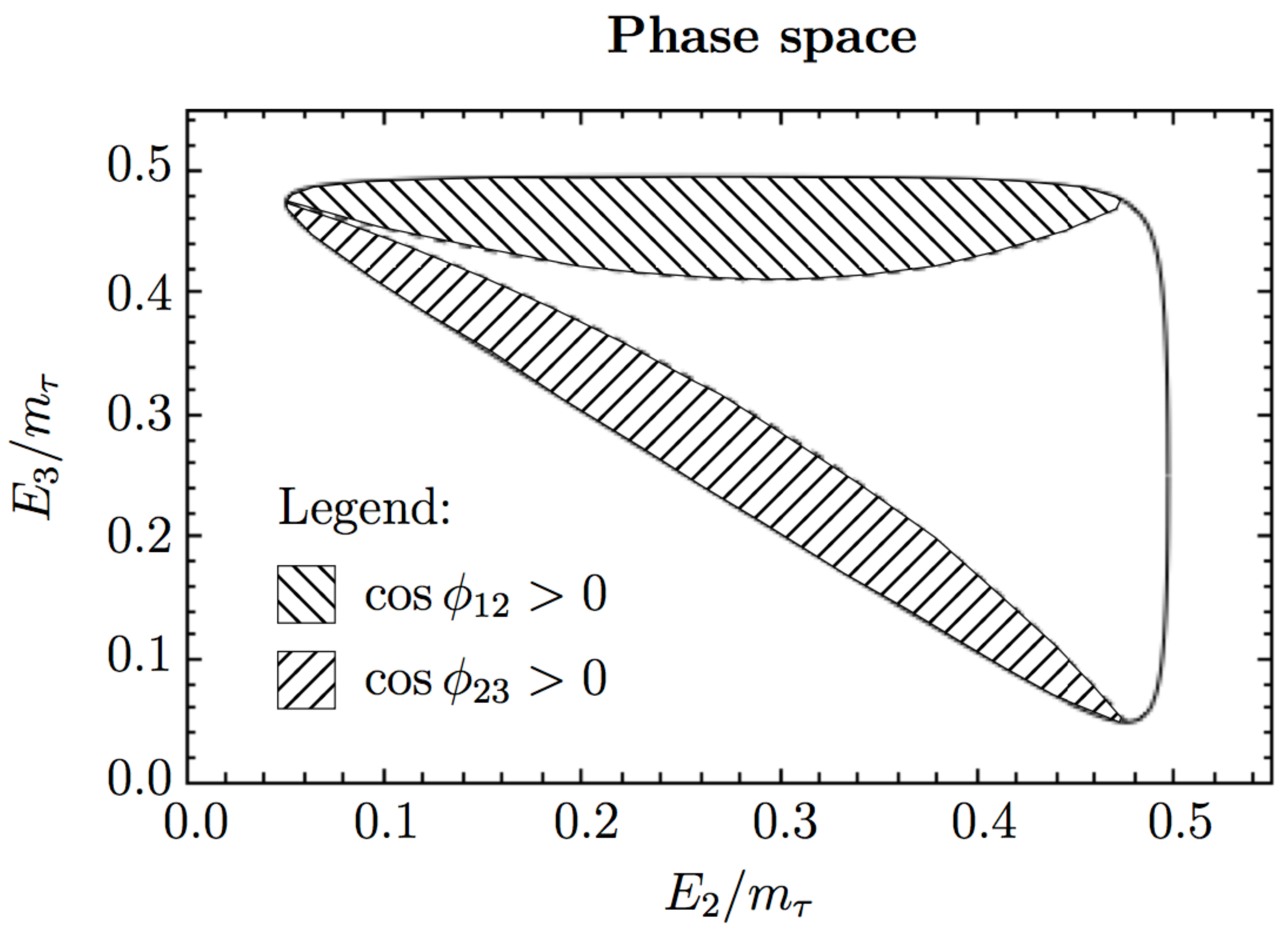, width=0.6\textwidth}
\caption{\label{fig:1}Available range of the energies $E_2$ and
  $E_3$. The shaded areas show the regions in which $\cos\phi_{23}>0$
  and where $\cos\phi_{12}>0$.}
\end{center}
\end{figure}

Besides the angles in (\ref{eq:15}) we choose two of the three
energies of the final-state leptons, which in the rest frame of the
tau satisfy $E_1 + E_2 + E_3 = m_\tau$. It is a straight-forward
exercise to show that the totally differential decay rate is then
given by
\begin{equation}
  \label{eq:19}
  \frac{d^5\Gamma(\tau^\uparrow \to 3 \ell)}{d\alpha\, d\!\cos\beta\,
    d\gamma\,dE_2\, dE_3} = \frac{1}{2m_\tau} \frac{|\mathcal
    M^\uparrow|^2}{256 \pi^5} \;.
\end{equation}
Since $|\mathcal M^\uparrow|^2$ does not depend on the angle $\gamma$
one can trivially integrate over the allowed range. The phase space
for the energies $E_2$ and $E_3$ is nearly triangular with edges that
are smoothened out by the light-lepton mass $m$. One way of expressing the
phase-space boundaries is, for example, 
\begin{equation}
  \label{eq:20}
  \frac12(m_\tau -E_2 - b_2 d_2)  \le E_3 \, \le \,  \frac12(m_\tau -E_2 +
 b_2 d_2) \comma
  m  \, \le \, E_2 \le  \, \frac{m_\tau^2-3m^2}{2m_\tau} \;,
\end{equation}
where $b_2 = \sqrt{E_2^2 - m^2}$ as before, and 
\begin{equation}
  \label{eq:21}
  d_2 = \frac{m_\tau^2-2m_\tau E_2 - 3m^2}{m_\tau^2-2m_\tau E_2 +m^2}
\end{equation}
is a function of $E_2$ that is near 1, except when $E_2$ is near the
endpoint where $d_2$ rapidly falls to zero. Sometimes it is more
advantageous to express these energies in Dalitz-like invariants
$s_{ij} = (p_i + p_j)^2$, for example when addressing the momentum
flowing through the virtual photon in the radiative operators in
(\ref{eq:2}). In terms of $s_{ij}$ the energies are
\begin{equation}
  \label{eq:22}
  E_2 = \frac{m_\tau^2 + m^2 - s_{13}}{2m_\tau} \comma 
  E_3 = \frac{m_\tau^2 + m^2 - s_{12}}{2m_\tau} \;.
\end{equation}
For later reference we also state the phase-space region in which
$\cos\phi_{23} \geq 0$, which is
\begin{eqnarray}
  \label{eq:23}
  \frac12(m_\tau -E_2 - b_2 d_2)  &\le E_3 \le &  
     \frac{m_\tau^2-2m_\tau E_2+m^2}{2(m_\tau-E_2)}  \;, \nonumber \\ 
  m &\le E_2 \le & \frac12 (m_\tau-m) \;.
\end{eqnarray}
Similarly, for $\cos\phi_{12} \geq 0$, we have the restrictions 
\begin{eqnarray}
  \label{eq:24}
 \frac{m_\tau(m_\tau-2E_2)+2E_2^2-m^2}{2(m_\tau - E_2)} &\le E_3 \le &  
     \frac12(m_\tau -E_2 + b_2 d_2)  \;, \nonumber \\ 
  m &\le E_2 \le & \frac12 (m_\tau-m) \;.
\end{eqnarray}
In Fig.~\ref{fig:1} we show the phase space for all kinematically
allowed values of $E_2$ and $E_3$ mentioned in equation~(\ref{eq:20})
as well as for the regions defined by (\ref{eq:23}) and (\ref{eq:24}).

\section{\label{sec:3}Definition of the Observables}

In this section we will use the formulae from the last chapter to
define observables, which can be measured even with sparse data
samples. All the observables will be defined from the coarsely sliced
phase space using the angular variables, so that the observables
correspond to partial rates and forward-backward asymmetries with
respect to the angles, which also involve the direction of the
tau-lepton polarization.
 
\subsection{\label{sec:3.1}Coupling bilinears}
Since lepton-flavour violating processes are rare, we focus on
observables where the available phase space in $\alpha, \beta, E_2,
E_3$ is at least partially integrated. There are many such
observables, and a measurement of them will allow us to draw
conclusions on the underlying operator structures. We start by
considering the integration of $J_i$ over the energies $E_2$ and $E_3$
in some region ``R'' of the phase space. The results are bilinears in
the couplings $\xi$ in (\ref{eq:4}), and can be expressed in terms of $6\times
6$ hermitian matrices $\bm{A}_i^{\rm (R)}$ that depend on the region R, to wit
\begin{equation}
  \label{eq:25}
\iint\limits_{\rm R} dE_2 \, dE_3\, J_i(E_2,E_3) = \frac{m_\tau^6}{6} \,
\xi^\dagger \, \bm{A}_i^{\rm (R)} \, \xi\;.
\end{equation}
From this notation it is obvious that the doubly differential
decay rate in $\alpha, \beta$ is obtained by integration over the full
energy phase space (\ref{eq:20}),
\begin{equation}
  \label{eq:26}
\frac{d^2\Gamma(\tau^\uparrow \to 3 \ell)}{d\alpha\, d\!\cos\beta}  =
\frac{m_\tau^5}{6\cdot256 \pi^4} \xi^\dagger \Big[ 
\bm{A}_1^{\rm \{full\}} + \bm{A}_2^{\rm \{full\}} \cos\beta + \bm{A}_3^{\rm \{full\}} \cos\alpha \sin\beta +
\bm{A}_4^{\rm \{full\}} \sin\alpha \sin\beta \Big] \xi\;,
\end{equation}
from which asymmetries using $\alpha, \beta$ are readily
calculable. For example, the difference in the partial rates
$\Gamma(\cos\beta > 0) - \Gamma(\cos\beta < 0)$ will only involve
$\bm{A}_2^{\rm \{full\}}$. We may express this difference as the
convolution integral over the differential decay rate (\ref{eq:26})
with the weight function $[\theta(\cos\beta) - \theta(-\cos\beta)]$,
where $\theta$ is the Heaviside step function. In order to demonstrate
that one can pick out each individual matrix using partial rates we
note that
\begin{equation}
  \label{eq:27}
  \int\limits_0^\pi d\alpha \int\limits_{-1}^{1}
  d\!\cos\beta \, \frac{d^2\Gamma}{d\alpha\,
    d\!\cos\beta} \, W(\alpha,\cos\beta) = \frac{m_\tau^5}{6\cdot256
    \pi^3} \xi^\dagger \left[ \sum\limits_{i=1}^4 \tilde c_i \bm{A}_i^{\rm \{full\}} \right] \xi\;,
\end{equation}
and list the coefficients $c_i$ for a few typical weight functions in
Table~\ref{tab:weights}. (Note the factor of $\pi$ that has been
absorbed into the prefactor.)
\begin{table}[t!]
\caption{\label{tab:weights}Coefficients of the linear combination in (\ref{eq:27}) for a
few typical weight functions. Since the matrices $\bm{A}_i^{\rm (R)}$
depend on the region R in the energy phase space we leave the upper
index $c$ on the partial rates open.}
\begin{center}
\begin{tabular}{l|c|cccc}
Partial rates combination & $W(\alpha,\cos\beta)$ 
& $\tilde c_1$ & $\tilde c_2$ & $\tilde c_3$ & $\tilde c_4$ \\ \hline
$\sum_{a,b}\Gamma_{ab}^c$ & 1 & 2 & 0 & 0 & 1 \\
$\sum_a (\Gamma_{a1}^c - \Gamma_{a2}^c) $ & $\theta(\cos\beta) - \theta(-\cos\beta)$ & 0 & 1 & 0 & 0 \\
$\sum_b (\Gamma_{1b}^c + \Gamma_{2b}^c - \Gamma_{3b}^c -
\Gamma_{4b}^c)$ 
& $\theta(\frac{\pi}{2}-\alpha) - \theta(\alpha - \frac{\pi}{2})$ & 0 & 0 & 1 & 0 \\
$\sum_b 2(\Gamma_{2b}^c+\Gamma_{3b}^c)$ 
& $2\theta(\frac{3\pi}{4}- \alpha) \theta(\alpha - \frac{\pi}{4})$ & 2
& 0 & 0 & $\sqrt{2}$
\end{tabular}
\end{center}
\end{table}
We define the following partial rates $\Gamma_{ab}^c$, which will form the basis of our
simulated counting experiments in Section~\ref{sec:4}, obtained from disjoint
regions of the phase space in $\alpha$, $\beta$, $(E_2,E_3)$. The
assignments are
\begin{equation}
\label{eq:GammaABC}
\Gamma_{ab}^c : \qquad
\begin{tabular}{c|c||c|c||c|c}
$a$ & $\alpha \in$ & $b$ & $\beta \in$ & $c$ & R \\ \hline
1 & $[0,\frac{\pi}{4})$ & 1 & $[0,\frac{\pi}{2})$ & 1 & $\{\cos\phi_{23}>0\}$ \\
2 & $[\frac{\pi}{4},\frac{\pi}{2})$ & 2 & $[\frac{\pi}{2}, \pi)$ & 2 &
$\{\cos\phi_{23}<0 \land \cos\phi_{12}<0 \}$ \\
3 & $[\frac{\pi}{2},\frac{3\pi}{4})$ & -- & -- & 3 & $\{\cos\phi_{12}>0 \}$ \\
4 & $[\frac{3\pi}{4}, \pi)$ & -- & -- & -- & --
\end{tabular} 
\end{equation}
Splitting the range of $\alpha$ into more than two regions is
necessary for isolating $\bm{A}_4^{\rm (R)}$.  Table~\ref{tab:weights}
also states the linear combination of these rates that correspond to
the given binning functions.  It is clear that {\it any} observable
constructed from the doubly differential decay rate (\ref{eq:26})
corresponds to a matrix that is such a linear combination of the four
$\bm{A}_i^{\rm \{full\}}$ matrices. In fact only two of these
structures contain information, since $\bm{A}_3^{\rm \{full\}} =
\bm{A}_4^{\rm \{full\}} =0$ due to the fact that the two final-state
leptons are identical particles and that $J_3$ and $J_4$ are
antisymmetric in the interchange of $E_1 \leftrightarrow E_2$. In
other words, any weighted integral over (\ref{eq:26}) is just a linear
combination of the total decay rate and the above-mentioned asymmetry.
Note also that $\bm{A}_4^{\rm \{full\}} =0$ is a necessary condition
for our discussion on the {\it unpolarized} decay rate around
equation~(\ref{eq:8}) as is evident from the first line in
Table~\ref{tab:weights}.
\\

If, however, we consider only the part of the phase space
(\ref{eq:23}) in which the angle between $\vec p_2$ and $\vec p_3$ is
between 0 and $\pi/2$ (denoted by R$=\{\cos \phi_{23}> 0\}$), all four
matrices $\bm{A}_i^{(\cos \phi_{23}> 0)}$ are non-zero and contribute
new information. (Note that R$=\{\cos \phi_{23}< 0\}$, which is the
complement to the full phase space, would then not yield more
information.) Similarly the angle between the two leptons carrying
momenta $\vec p_1$ and $\vec p_2$ can be utilized. The region
R$=\{\cos \phi_{12}> 0\}$ contributes two more non-zero matrices as
again $\bm{A}_3^{(\cos \phi_{12}> 0)} = \bm{A}_4^{(\cos \phi_{12}> 0)}
= 0$. We therefore count eight
different partial rates from which to construct observables. \\

Since the vector of coupling constants, $\vec \xi$, 
contains six complex parameters, i.e.\ twelve real unknowns,
the above partial rates do not suffice to solve the system. 
One way out would be to obtain independent information on the 
coefficients $\xi_5$ and $\xi_6$ from the radiative decays $\tau \to \mu\gamma$.
Alternatively, one could, of course, divide the phase space into more
(i.e.\ smaller) regions $R$, but this would only make sense if sufficient
signal events had been measured. For the time being, we will restrict
ourselves to a handful of benchmark scenarios that will be defined
and analyzed in Sec.~\ref{sec:4}.

\subsection{Calculation of the matrices $\bm{A}_i^{\rm (R)}$}

The tree-level calculation of the squared amplitude is straightforward,
and we have collected the various parts in
Appendix~\ref{apx:ME}. The mass $m$ of the light leptons in the final
state has been kept finite in our calculations.
The task is now to integrate the resulting
functions $J_i(E_2,E_3)$ over (part of) the phase space $R$ to arrive at
the matrices $\bm{A}_i^{\rm (R)}$. Their entries are functions of the
mass ratio ratio $\ee = m/m_\tau$, which is small. It is tempting
to state the results analytically in an expansion of $\ee$, which can
be done using the method of regions \cite{Beneke:1997zp}. We
give a brief discussion of this strategy on two illustrative examples
which can be found in Appendix~\ref{apx:MR}.
For practical purposes, it suffices to evaluate the matrices numerically,
and we list the results in Appendix~\ref{apx:2} for both $\tau \to 3
\mu$ and $\tau \to 3 e$ decays.

\subsection{\label{sec:litcomp}Comparison with the literature}

The proposal of utilising the polarization vector of the tau for
asymmetries \cite{Mannel:2014sba} was preceded by an analysis within
the context of the littlest Higgs model with $T$ parity
\cite{Goto:2010sn}, in which the authors also considered lepton-flavour
violating decays of polarized (anti-) taus into light leptons, among
other channels. Contrary to our assumption stated after equation
(\ref{eq:1}) that $(LR)(LR)$ and $(RL)(RL)$ operators are suppressed
by $(v^2/\Lambda^2)$, Goto et~al.~included them in their work. 
Furthermore only the leading-power expressions in $(m/m_\tau)$ are
considered -- massless final-state leptons, in other words. With this
approximation there is no interference between operators of different
chirality, which require a mass insertion. In this limit the couplings
of the above-mentioned extra operators, called $g_{Ls}^{\rm I}$ and
$g_{Rs}^{\rm I}$ enter simply by adding them to $g_V^{(LL)(LL)}$ and
$g_V^{(RR)(RR)}$ (in our notation) in quadrature, see (A8c) of
\cite{Goto:2010sn}. \\

When neglecting final-state lepton masses one encounters unregulated
collinear and soft divergences in the contribution from the radiative
operators at the boundaries of phase space, where the intermediate
photon propagator can become on-shell. Goto et~al.~introduced a cutoff
parameter $\delta>0$ in order to stay away from this boundary, so that
their expressions become functions of $\delta$ rather than
$(m/m_\tau)$. Here, in this present work, we went further in that the
final-state lepton masses remain finite, and therefore all
interference terms are accounted for and all soft and collinear
divergences are naturally regulated. \\

We have compared some of the terms in the fully differential decay
rate stated in \cite{Goto:2010sn} with the corresponding leading-power
approximations of our results in Appendix~\ref{apx:ME} and find
agreement after accounting for the different setups. 
\\

\section{\label{sec:4}Phenomenology}

We start with assuming some values for the LFV couplings, on the basis of which 
we generate $\mathcal{N}_{\rm total}$ events and bin them into $\mathcal
N_{ab}^c$ counts congruent with the definition of the partial rates in
(\ref{eq:GammaABC}). Our goal is then to reconstruct the couplings
from a simple and straight-forward least-square fit of the binning
counts, i.e.~fitting the bin probabilities,
\begin{equation}
  \label{eq:60}
  \prob_{ab}^c(\xi) = \frac{\Gamma_{ab}^c}{\Gamma} = \frac{\xi^\dagger
    \bm{A}_{ab}^c \xi}{\xi^\dagger \bm{A}_1^{\rm \{ full \}} \xi} \,,
\end{equation}
to the fraction of events in that bin, $\mathcal
N_{ab}^c/\mathcal{N}_{\rm total}$. Note that the bin probabilities are
invariant under a simultaneous rescaling of all couplings, $\xi
\longrightarrow \omega\xi$. In order to avoid this flat direction in
the $\chi^2$ function we impose a condition that breaks the rescaling
invariance, to wit $\xi^\dagger \bm{A}_1^{\rm \{ full \}} \xi \equiv 
1$ . We stress that a thus fitted result only reflects the relative
strength of the couplings to each other, and not the values of the
couplings themselves. Those can be inferred from the total decay width
of this process and the lifetime of
the tau, which is not our main focus here.\footnote{We stress however,
that a measurement of the total decay width itself
does depend on the underlying distribution 
over the energy and angular phase space, and therefore -- in case of only a few events --
requires the information about the relative size of the individual couplings 
in an essential way. This also affects the experimental procedure to
generate bounds on $\Gamma(\tau \to 3 \ell)$.}
\\

As we have seen in Section~\ref{sec:3.1} there are eight independent
matrices from which we can calculate observables. One of them,
$\bm A_4^{\{\cos\phi_{23}>0\}}$, probes the imaginary parts of
the couplings and does not contribute if the couplings are real. One
may choose many different sets of observables to fit for the
couplings, but here we simply use the bins from which these
observables are calculated themselves. The least-square fit is thus
performed by minimizing the function
\begin{equation}
  \label{eq:61}
  \chi^2(\xi,\mu) = \sum\limits_{a,b,c} \left( \frac{\mathcal N_{ab}^c
    - \xi^\dagger \bm{A}_{ab}^c \xi \,\mathcal{N}_{\rm total} }{\Delta 
    \mathcal N_{ab}^c}  \right)^2 + 2 \mu \left( 
  \xi^\dagger \bm{A}_1^{\rm \{ full \}} \xi -1 \right)\;,
\end{equation}
where $\mu$ is the Lagrange multiplier used to impose the above
normalization condition. We use the unsophisticated statistical
estimator for the uncertainty of each bin count $\Delta \mathcal
N_{ab}^c = \sqrt{\mathcal N_{ab}^c}$ if $\mathcal N_{ab}^c \ne 0$ and
1 otherwise.\\

In the following, as a premature study, 
we are assuming only {\it real} couplings, for simplicity.
In this case the above-mentioned matrix $\bm
A_4^{\{\cos\phi_{23}>0\}}$ does not contribute, and there is no need
for splitting the range of $\alpha$ into four distinct bins. Hence we
will combine the bins with $a=1$ and $a=2$ into one single bin, as
well as the bins $a=3$ and $a=4$. We will therefore fit 12 observables
(seven of which are independent)
to 6 real unknowns.

\subsection{Muonic final state}

\begin{figure}[t!b!!p]
\centering
\epsfig{file=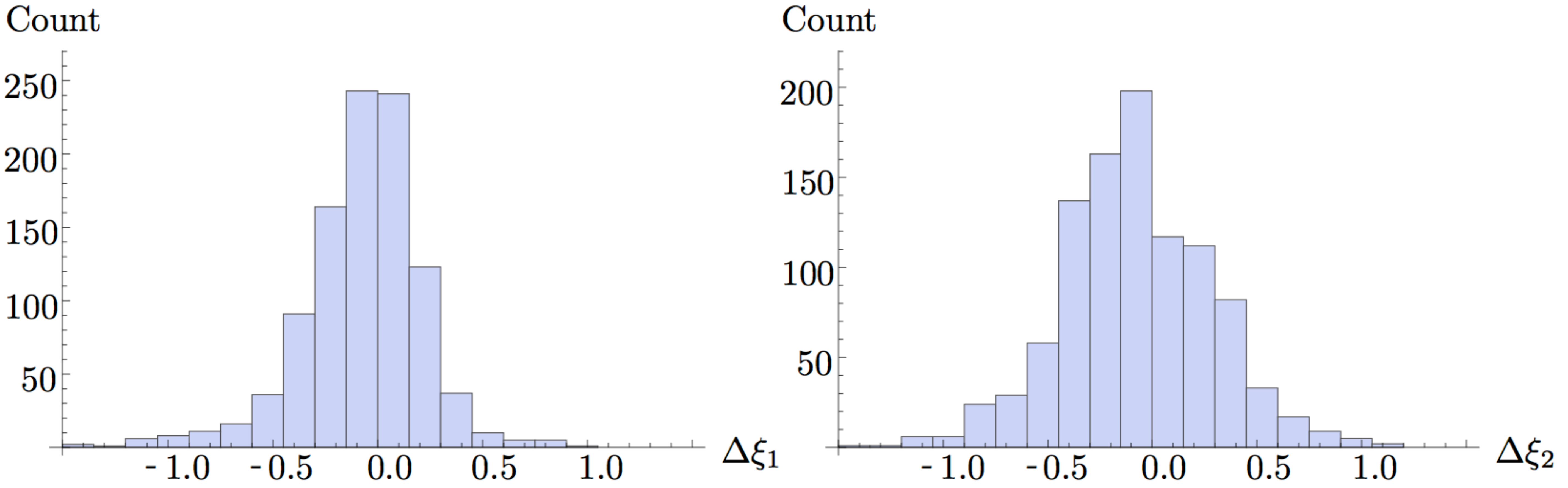, width=150mm}
\caption{\label{fig:2}Example histograms of the difference $\Delta
  \xi_1 = \xi_1^{\rm fit} - \xi_1^{\rm in}$ (left) and $\Delta \xi_2$
  (right) for $\mathcal{N}_{\rm total} = 100$ events in scenario
  a). The standard deviation of the {\bf left} distribution is
  $\sigma_1 = 0.288$, while the mean of the individual fit errors is
  $\overline{\delta \xi_1^{\rm fit}} = 0.256$. This example shows a
  shape that is more peaked than a normal distribution. The {\bf right}
  distribution has $\sigma_2 = 0.362$ which is less than $\overline{\delta
    \xi_1^{\rm fit}} =0.421$, and the shape is more box-like than a normal
distribution.}
\end{figure}

For a first impression we have randomly chosen a scenario
with real couplings:
\begin{eqnarray}
  \label{eq:62}
  \hbox{Scenario a) } &\qquad & 
  \xi^T = \omega\, (1.3,0.4,-4.0,0.0,2.4,1.2) \nonumber \\
  && \xi^\dagger \bm{A}_1^{\rm \{ full \}} \xi = 1 \qquad \Rightarrow
  \qquad \omega = 0.1086 \;.
\end{eqnarray}
A typical entry in $\xi$ is therefore $\mathcal O(0.1)$ to which we may compare
the errors. From the probabilities we generate $\mathcal N_{\rm
  total}$ events that are distributed in the bin counts $\mathcal
N_{ab}^c$. These serve as the output of our virtual
experiment. Notice that in this example, all couplings have entries 
of similar size (or happen to be zero). 
At this point we pretend that the couplings $\xi$ are unknown and
proceed with the fit. The outcome $\xi^{\rm fit}$ will in general
deviate from the scenario input $\xi^{\rm in}$, and we may form the
deviation vector $\Delta \xi = \xi^{\rm fit} - \xi^{\rm in}$, not to
be confused with the individual fit errors $\delta \xi^{\rm fit}$. We
then repeat this virtual experiment one thousand times and display the
entries of the deviation vector $\Delta \xi$ in histograms, which peak
around zero with a certain width (see Fig.~\ref{fig:2}). If a
histogram shows a normal distribution then the width ($2\sigma$)
coincides with twice the mean of the fit errors $\overline{\delta
  \xi^{\rm fit}}$.
\begin{table}[t!bp]
  \caption{\label{tab:2}Histogram widths $2\sigma_i$ as the error estimator for samples containing
    $\mathcal{N}_{\rm total}$ events, as well as the mean fit
    uncertainty for Scenario a). }
  \begin{center}  
  \begin{tabular}{c|cccccc}
    $\mathcal{N}_{\rm total}$ 
    & $(\sigma_1,\overline{\delta \xi^{\rm fit}_1})$ 
    & $(\sigma_2,\overline{\delta \xi^{\rm fit}_2})$ 
    & $(\sigma_3,\overline{\delta \xi^{\rm fit}_3})$ 
    & $(\sigma_4,\overline{\delta \xi^{\rm fit}_4})$ 
    & $(\sigma_5,\overline{\delta \xi^{\rm fit}_5})$ 
    & $(\sigma_6,\overline{\delta \xi^{\rm fit}_6})$
    \\ \hline
    20 & (0.51, 0.51) & (0.51, 0.69) & (0.47, 0.85) & (0.50, 0.88) &
    (0.15, 0.14) & (0.17, 0.19) \\
    100 & (0.29, 0.26) & (0.36, 0.42) & (0.31, 0.43) & (0.36, 0.48) &
    (0.09, 0.05) & (0.14, 0.12) \\
    1000 & (0.09, 0.10) & (0.18, 0.19) & (0.18, 0.22) & (0.17, 0.24) &
    (0.02, 0.02) & (0.03, 0.04) \\
    10000 & (0.03, 0.04) & (0.05, 0.06) & (0.12, 0.11) & (0.09, 0.09)
    & (0.01, 0.01) & (0.01, 0.01)
  \end{tabular}
  \end{center}
\end{table}  
In Table~\ref{tab:2} the resulting widths of the histograms
$( = 2\sigma_i)$ are listed for a few numbers of events $\mathcal N_{\rm
  total}$. We calculate $\sigma_i$ as the standard deviation of the
data underlying the histogram. The reader should keep in mind that the input values of the
couplings are $\xi^T_{\rm in} \approx (0.14, 0.04, -0.43, 0.00, 0.26, 0.13)$
which is to be compared to the uncertainty of the fit results.
Regardless of whether we interpret $\sigma_i$ or
$\overline{\delta \xi_i^{\rm fit}}$ as the typical uncertainty of the
coupling $\xi_i^{\rm fit}$, the conclusion is that with a few tens of events
  only the couplings of the radiative operators, $\xi_5$ and $\xi_6$,
  can be extracted in any meaningful way, while the couplings of the
  leptonic operators, $\xi_1$ through $\xi_4$ require hundreds -- if
  not thousands -- of events.  This pattern is a direct consequence
of the magnitude of the entries in the $\bm{A}_i^{\rm (R)}$ matrices,
see e.g.~(\ref{eq:42}): The largest entries are the log-enhanced
diagonal elements of the radiative sector, which is why the
observables are quite sensitive to $\xi_5$ and
$\xi_6$ (unless these couplings happen to be generically suppressed in 
 a particular class of NP models under consideration). 
\\

\begin{figure}[t!p!]
  \centering
  \epsfig{file=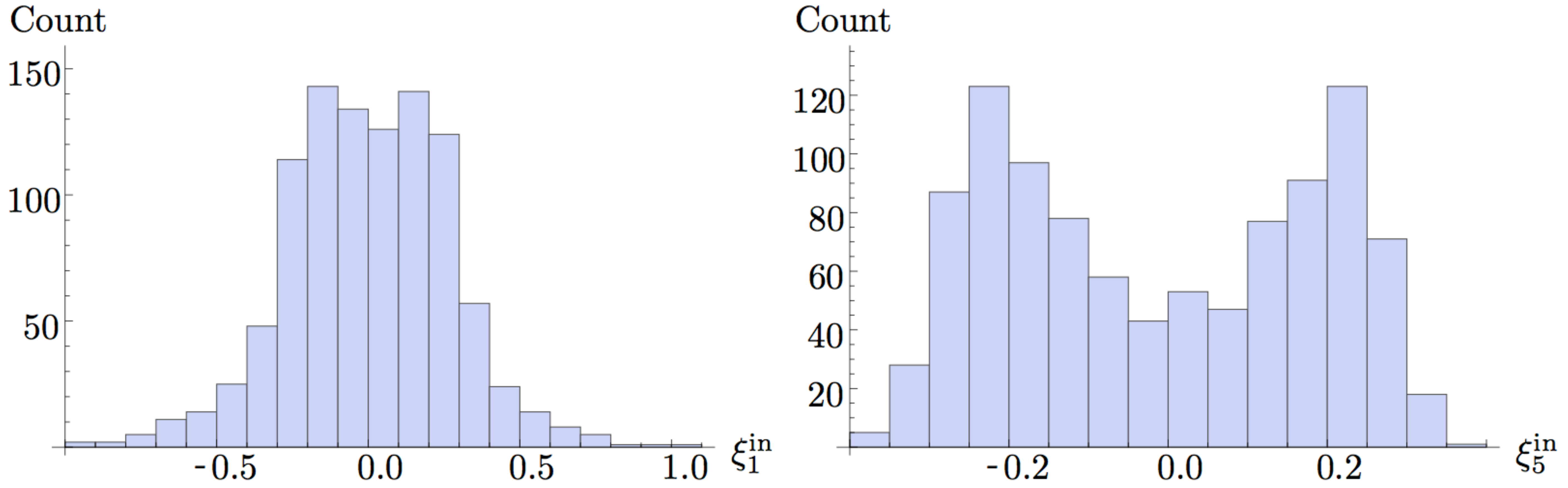, width=150mm}
  \caption{\label{fig:3}Distributions of randomly chosen couplings,
    rescaled for $\xi^\dagger \bm{A}_1^{\rm \{ full \}} \xi = 1$,
    i.e.~Scenario~b). Shown here are $\xi_1^{\rm in}$ (left) and
    $\xi_5^{\rm in}$ (right). The remaining couplings of leptonic
    operators, $\xi_2^{\rm in}$ through $\xi_4^{\rm in}$, are
    distributed similar to $\xi_1^{\rm in}$, and the second radiative
    coupling, $\xi_6^{\rm in}$, similar to $\xi_5^{\rm in}$. The mean
    absolute values of any coupling is around $0.2$. }
\end{figure}

One may ask if the particular choices of parameter values in scenario a) have any influence
on the outcome of this sensitivity study. We therefore repeat the
above procedure with the following twist:
\begin{eqnarray}
  \label{eq:63}
  \hbox{Scenario b)} &\qquad & \hbox{For each of the one thousand
    samples the couplings are chosen at random } \nonumber \\
&& \hbox{and rescaled to abide
    by} \quad \xi^\dagger \bm{A}_1^{\rm \{ full \}} \xi = 1 \;.
\end{eqnarray}
We generate a random set of couplings, drawn from a finite
interval, which we choose symmetric around zero and universal for all 
couplings $\xi_i$, and rescale to build a sample of vectors
$\vec \xi$. This procedure removes the preference towards particular
values for the couplings, although implicitly it assumes that they 
are still of the same order of magnitude. 
Notice that, as a consequence for the condition
(\ref{eq:63}), the resulting distributions in the
sampled couplings $\xi_i$ are not flat. In Figure~\ref{fig:3} we show the
distributions for $\xi_1$ and $\xi_5$, exemplary for couplings to
leptonic and radiative operators, respectively. Note that the means
for the absolute values are roughly around $0.2$ in both cases, in
accordance with our previous observation that couplings are of order
$\mathcal O(0.1)$. 
We present the corresponding fit results in Table~\ref{tab:3}, which may be
juxtaposed to the previous scenario. In general the widths $\sigma_i$
have increased due to the fact that one is convoluting the
``fixed-coupling results'' with the ``coupling distributions'',
leading to more pronounced shoulders in the resulting
distributions. However, the mean fit uncertainties $\overline{\delta
\xi^{\rm fit}_i}$ are roughly the same, with some entries larger than
their fixed-coupling counterparts in Table~\ref{tab:2}, and some
entries smaller. Our general conclusion remains unchanged.

\begin{table}[t!bp!!]
  \caption{\label{tab:3}Histogram widths $2\sigma_i$ as the error estimator for samples containing
    $\mathcal{N}_{\rm total}$ events, as well as the mean fit
    uncertainty for Scenario b). } 
  \begin{center}
  \begin{tabular}{c|cccccc}
    $\mathcal{N}_{\rm total}$ 
    & $(\sigma_1,\overline{\delta \xi^{\rm fit}_1})$ 
    & $(\sigma_2,\overline{\delta \xi^{\rm fit}_2})$ 
    & $(\sigma_3,\overline{\delta \xi^{\rm fit}_3})$ 
    & $(\sigma_4,\overline{\delta \xi^{\rm fit}_4})$ 
    & $(\sigma_5,\overline{\delta \xi^{\rm fit}_5})$ 
    & $(\sigma_6,\overline{\delta \xi^{\rm fit}_6})$
    \\ \hline
    20 & (0.52, 0.59) & (0.53, 0.62) & (0.57, 0.82) & (0.59, 0.82) &
    (0.17, 0.16) & (0.18, 0.17) \\
    100 & (0.40, 0.35) & (0.41, 0.37) & (0.45, 0.43) & (0.48, 0.43) &
    (0.12, 0.08) & (0.14, 0.09) \\
    1000 & (0.22, 0.18) & (0.23, 0.18) & (0.24, 0.18) & (0.22, 0.18) &
    (0.06, 0.04) & (0.06, 0.04) \\
    10000 & (0.09, 0.07) & (0.11, 0.07) & (0.09, 0.07) & (0.10, 0.07)
    & (0.02, 0.01) & (0.02, 0.01)
  \end{tabular}
  \end{center}
\end{table}

We stress again that even with these randomized couplings for each
virtual experiment there is still a build-in assumption: that all
couplings are of the same order of magnitude with a flat distribution.
However, depending on the explicit New Physics model (see
e.g.~\cite{Ellis:2002fe,Brignole:2004ah,Paradisi:2005tk,Blanke:2007db,Akeroyd:2009nu,Buras:2010cp})
the relative importance of radiative and 4-lepton operators can be
quite different.  We therefore also considered a scenario in which the
radiative operators may be loop induced and are therefore accompanied by
Wilson coefficients that could be much smaller compared to those of the
leptonic operators. We may name this Scenario b'). To be concrete,
here we assume a flat distribution for the couplings of the radiative
operators that have a smaller support interval, by a factor of
$1/(4\pi)^2$. Again the absolute errors quoted in Table~\ref{tab:3}
remain the same, up to small variations. However, the corresponding
distributions in the spirit of Figure~\ref{fig:3} are such that
$\overline{\xi_1^{\rm in}}$ is of $\mathcal O(1)$ -- representativ for
all couplings to leptonic operators -- and the ones to radiative
operators are of $\mathcal O(1/100)$. In this case it is the leptonic
couplings that can be determined with a few dozen events, while the
radiative ones are elusive.  If this pattern were to be observed, we
could also expect a suppression of the $\tau \to \ell \gamma$ decay
channel. Ultimately a global fit with all relevant decay channels
would be in order.

\paragraph{Using Asymmetries to distinguish two Scenarios.}

In this subsection of the analysis the goal is to analyze the
discriminating power of our approach to distinguish two
different benchmark scenarios that may reflect the
dynamical effects of some classes of NP models.
Let us say that the dimension-6 operators (\ref{eq:3})
are induced by parity-violating interactions, i.e.\ one that couples
only to left-handed taus or only to right-handed taus,
\begin{eqnarray}
  \label{eq:53}
    \hbox{Scenario c)} &\qquad & \hbox{Only left-handed taus
      participate} \nonumber \\
    && \xi = \omega\,(1,0,1,0,0,1)\qquad , \qquad \xi^\dagger
    \bm{A}_1^{\rm \{ full \}} \xi = 1 \;, \\
    \hbox{Scenario d)} &\qquad & \hbox{Only right-handed taus
      participate} \nonumber \\
    && \xi = \omega\,(0,1,0,1,1,0)\qquad , \qquad \xi^\dagger
    \bm{A}_1^{\rm \{ full \}} \xi = 1 \;.
\end{eqnarray} Specifically
and for simplicity we consider in scenario~c) equal coupling constants
to all operators involving left-handed taus, while all other couplings
vanish. Scenario d) is the equivalent setup with right-handed
taus. \\

\begin{table}[t!bp]
  \centering
  \caption{\label{tab:4}Central values and standard deviations for
    the asymmetry $\mathcal A_\beta$ in both scenario c) and d), using
  1000 virtual experiments.}
  \begin{tabular}{c|c|c|c}
    $\mathcal N_{\rm total}$ & $\mathcal A_\beta$(Scen. c) & $\mathcal
    A_\beta$(Scen. d) &  Degree of Separation \\ \hline
      5 & $0.20 \pm 0.44$ & $-0.21 \pm 0.45$ & $0.36$ \\
    10 & $0.19 \pm 0.30$ & $-0.20 \pm 0.32$ & $0.47$ \\ 
    20 & $0.21 \pm 0.22$ & $-0.21 \pm 0.21$ & $0.67$ \\ 
    50 & $0.20 \pm 0.14$ & $-0.20 \pm 0.14$ & $0.85$ \\ 
  100 & $0.20 \pm 0.10$ & $-0.20 \pm 0.10$ & $0.95$ 
  \end{tabular}
\end{table}

There are several asymmetries one can construct from the angles
$\alpha, \beta, \phi_{12}, \phi_{23}$. Some of them have to be
combined, for example the asymmetry in $\cos \alpha$: according to
Table~\ref{tab:weights} it is proportional to $\bm{A}_3^{\rm (R)}$, but for
the full phase space $\bm{A}_3^{\rm \{full\}}=0$, so one needs to
further cut on the full phase space to gain information. 
Let us instead look at the asymmetry of the angle $\beta$, to wit
\begin{equation}
  \label{eq:65}
  \mathcal A_\beta = \frac{\Gamma(\cos\beta>0) - \Gamma(\cos\beta<0)}%
  {\Gamma(\cos\beta>0) + \Gamma(\cos\beta<0)}
  = \frac{1}{\mathcal N_{\rm total}} \sum\limits_{a,c} 
  (\mathcal N_{a1}^c - \mathcal N_{a2}^c)\;.
\end{equation}
Given the above two models we can readily calculate the theoretical
expectations of $\mathcal A_\beta = 0.202$ (Scenario c) and $\mathcal
A_\beta = -0.202$ (Scenario d). Even with very few total events one
can indeed already obtain an impression which scenario to prefer, as
is shown in Table~\ref{tab:4}.  Here, again, we have repeated the
virtual experiment 1000 times to produce a distribution of results
from which we estimate the typical error as the standard deviation of
the distribution. The central values of the distributions fluctuate a
little around the theoretical expectations due to the finite number of
virtual experiments, but even with a rather small $\mathcal N_{\rm
  total}$ the likelihood for one scenario over the other is
significant.%
\footnote{Interestingly the situation is even better for the
  corresponding scenarios in which all radiative operators are absent,
  $\xi_5 = \xi_6 = 0$. The central values then shift to $\mathcal
  A_\beta = \pm 0.257$ and the errors remain the same, so that the
  separation between the two datasets becomes more pronounced.} The
``Degree of Separation'' between the two scenarios
is calculated from the overlap of two normal distributions,
$\rho_c(\mathcal A)$ and $\rho_c(\mathcal A)$, with the
given central values and standard deviations of scenario c) and d), respectively,
\begin{equation}
  \label{eq:69}
  \hbox{Degree of Separation} = 1 - \int\limits_{-\infty}^\infty
  d\mathcal A \, \hbox{Min}\left[\rho_c(\mathcal A),\rho_d(\mathcal A)
  \right] \;.
\end{equation}
With this definition the Degree of Separation is zero for completely
overlapping distributions and asymptotically approaching 1.0 for
distributions that are far apart.

\subsection{Electronic final state}

\begin{table}[b!]
  \centering
  \caption{\label{tab:5}Histogram widths $2\sigma_i$ as the error estimator for samples containing
    $\mathcal{N}_{\rm total}$ events, as well as the mean fit
    uncertainty for Scenario a). } 
  \begin{tabular}{c|cccccc}
    $\mathcal{N}_{\rm total}$ 
    & $(\sigma_1,\overline{\delta \xi^{\rm fit}_1})$ 
    & $(\sigma_2,\overline{\delta \xi^{\rm fit}_2})$ 
    & $(\sigma_3,\overline{\delta \xi^{\rm fit}_3})$ 
    & $(\sigma_4,\overline{\delta \xi^{\rm fit}_4})$ 
    & $(\sigma_5,\overline{\delta \xi^{\rm fit}_5})$ 
    & $(\sigma_6,\overline{\delta \xi^{\rm fit}_6})$
    \\ \hline
    20 & (0.33, 0.42) & (0.42, 0.57) & (0.31, 0.74) & (0.31, 0.86) &
    (0.055, 0.039) & (0.055, 0.067) \\
    100 & (0.24, 0.27) & (0.24, 0.37) & (0.14, 0.38) & (0.18, 0.42) &
    (0.014, 0.014) & (0.043, 0.037) \\
    1000 & (0.12, 0.12) & (0.14, 0.16) & (0.08, 0.17) & (0.11, 0.15) &
    (0.004, 0.004) & (0.009, 0.010) \\
    10000 & (0.05, 0.05) & (0.10, 0.08) & (0.09, 0.09) & (0.08, 0.07)
    & (0.002, 0.002) & (0.004, 0.004)
  \end{tabular}
  \caption{\label{tab:6}Histogram widths $2\sigma_i$ as the error estimator for samples containing
    $\mathcal{N}_{\rm total}$ events, as well as the mean fit
    uncertainty for Scenario b). } 
  \begin{tabular}{c|cccccc}
    $\mathcal{N}_{\rm total}$ 
    & $(\sigma_1,\overline{\delta \xi^{\rm fit}_1})$ 
    & $(\sigma_2,\overline{\delta \xi^{\rm fit}_2})$ 
    & $(\sigma_3,\overline{\delta \xi^{\rm fit}_3})$ 
    & $(\sigma_4,\overline{\delta \xi^{\rm fit}_4})$ 
    & $(\sigma_5,\overline{\delta \xi^{\rm fit}_5})$ 
    & $(\sigma_6,\overline{\delta \xi^{\rm fit}_6})$
    \\ \hline
    20 & (0.38, 0.48) & (0.40, 0.49) & (0.38, 0.75) & (0.37, 0.76) &
    (0.060, 0.049) & (0.066, 0.052) \\
    100 & (0.28, 0.31) & (0.29, 0.34) & (0.25, 0.38) & (0.25, 0.39) &
    (0.037, 0.023) & (0.043, 0.026) \\
    1000 & (0.18, 0.17) & (0.18, 0.17) & (0.16, 0.15) & (0.17, 0.16) &
    (0.010, 0.009) & (0.010, 0.09) \\
    10000 & (0.09, 0.07) & (0.09, 0.07) & (0.08, 0.06) & (0.07, 0.06)
    & (0.006, 0.004) & (0.005, 0.004)
  \end{tabular}
\end{table}

When the final-state leptons are electrons the operators and Wilson
coefficients are different and in general independent from the
above-mentioned decay into muons. Since electrons are much lighter
than muons the entries of the $\bm{A}_i^{\rm (R)}$ matrices change,
and the results can be found in Appendix~\ref{apx:2}. Again we
start our phenomenological game analogous to the muonic case above with
\begin{eqnarray}
  \label{eq:66}
    \hbox{Scenario a) } &\qquad & 
  \xi^T = \omega\, (1.3,0.4,-4.0,0.0,2.4,1.2) \nonumber \\
  && \xi^\dagger \bm{A}_1^{\rm \{ full \}} \xi = 1 \qquad \Rightarrow
  \qquad \omega = 0.0503 \;.
\end{eqnarray}
Although we assume here the same relative coupling strengths as in the
muonic example above, we remind the reader that there is no relation to
the muonic case, and the goal is simply to gain insights into our
overall ability to determine the couplings from $\mathcal N_{\rm
  total}$ events. However, the normalization condition leads to
typical coupling magnitudes that are about half as large as in the
muonic case, see (\ref{eq:62}). Comparing the results
in Table~\ref{tab:5} with those in Table~\ref{tab:2}, we observe that
the errors on the couplings of leptonic operators are somewhat smaller
for electronic final states, but not by half as is the case for the
couplings themselves. However, the radiative couplings are now far easier
to accurately determine. For example, with only $\mathcal N_{\rm
  total} = 20$ events the typical fit result yields $\xi_5 = 0.121 \pm
0.039$, whereas $\xi_1 = 0.065 \pm 0.42$ does not allow us much
insight.

\begin{eqnarray}
  \label{eq:67}
    \hbox{Scenario b)} &\qquad & \hbox{For each of the one thousand
    samples the couplings are chosen at random } \nonumber \\
&& \hbox{and rescaled to abide
    by} \quad \xi^\dagger \bm{A}_1^{\rm \{ full \}} \xi = 1 \;.
\end{eqnarray}

\begin{figure}[tp]
  \centering
  \epsfig{file=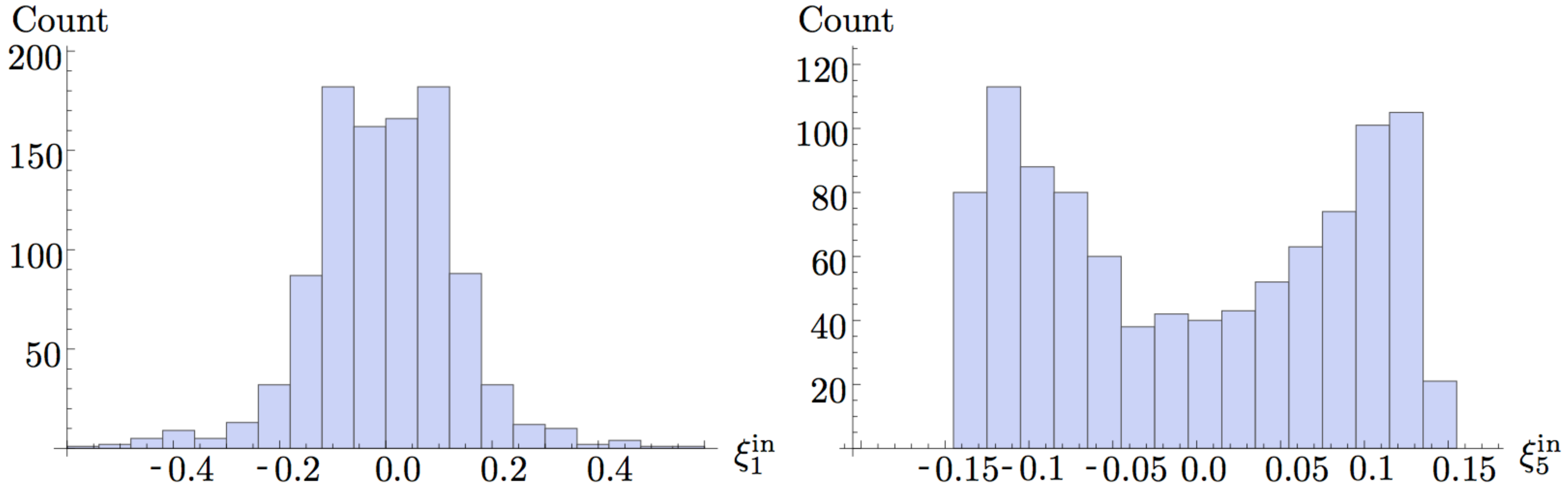, width=150mm}
  \caption{\label{fig:4}Distributions of randomly chosen couplings,
    rescaled for $\xi^\dagger \bm{A}_1^{\rm \{ full \}} \xi = 1$,
    i.e.~Scenario~b). Shown here are $\xi_1^{\rm in}$ (left) and
    $\xi_5^{\rm in}$ (right). The remaining couplings of leptonic
    operators, $\xi_2^{\rm in}$ through $\xi_4^{\rm in}$, are
    distributed similar to $\xi_1^{\rm in}$, and the second radiative
    coupling, $\xi_6^{\rm in}$, similar to $\xi_5^{\rm in}$. The mean
    absolute values of any coupling is around $0.1$.}
\end{figure}

In Figure~\ref{fig:4} we show the distribution obtained from
randomizing the couplings and rescaling. Again the typical size of
couplings is of order $\mathcal O(0.1)$. Just as in the previous case
we observe no significant difference from our finding with fixed
couplings (compare Table~\ref{tab:5} to Table~\ref{tab:6}). Next, we
look at the ability to distinguish two scenarios,
\begin{eqnarray}
  \label{eq:68}
      \hbox{Scenario c)} &\qquad & \hbox{Only left-handed taus
      participate} \nonumber \\
    && \xi = \omega\,(1,0,1,0,0,1)\qquad , \qquad \xi^\dagger
    \bm{A}_1^{\rm \{ full \}} \xi = 1 \;, \\
    \hbox{Scenario d)} &\qquad & \hbox{Only right-handed taus
      participate} \nonumber \\
    && \xi = \omega\,(0,1,0,1,1,0)\qquad , \qquad \xi^\dagger
    \bm{A}_1^{\rm \{ full \}} \xi = 1 \;.
\end{eqnarray}

\begin{table}[b]
  \centering
  \caption{\label{tab:7}Central values and standard deviations for
    the asymmetry $\mathcal A_\beta$ in both scenario c) and d), using
  1000 virtual experiments.}
  \begin{tabular}{c|c|c|c}
    $\mathcal N_{\rm total}$ & $\mathcal A_\beta$(Scen. c) & $\mathcal
    A_\beta$(Scen. d) & Degree of Separation \\ \hline
      5 & $0.40 \pm 0.413$ & $-0.39 \pm 0.412$ & $0.662$ \\
    10 & $0.40 \pm 0.280$ & $-0.37 \pm 0.298$ & $0.817$ \\
    20 & $0.39 \pm 0.214$ & $-0.39 \pm 0.204$ & $0.938$ \\
    50 & $0.39 \pm 0.131$ & $-0.38 \pm 0.129$ & $0.997$ \\
  100 & $0.39 \pm 0.091$ & $-0.39 \pm 0.095$ & $1.000$ 
  \end{tabular}
\end{table}

The central values for the asymmetry (\ref{eq:65}) in these two
scenarios is $\mathcal A_\beta = \pm 0.387$. We show the outcome of
1000 simulations for each $\mathcal N_{\rm total}$ in
Table~\ref{tab:7}. It is notable that even very few events suffice for
the distinction of these two scenarios.

\newpage
\section{\label{sec:5}Summary}

Since the decay $\tau \to 3\ell$ is very clean, already a single event
of this type would immediately imply New Physics, since the prediction
of the Standard Model (extended by including the neutrino masses
through the Weinberg operator) is practically zero. However, once such
a decay would be observed, the nature of the underlying interaction
has to be uncovered. In general this would require to study decay
distributions, which is impossible with only a few events.

In this paper we have studied the LFV decays of tau leptons into three
muons or electrons, including a polarization of the tau lepton. Such a
polarization can be generated from running an $e^+ e^-$ collider in
the vicinity of the $\tau^+ \tau^-$ threshold with polarized electrons
and/or positrons. Experimentally this could be realized at BES III,
once a polarization of the beams would become possible.

We have considered a general, model-independent set-up for the
interaction mediating the $\tau$ decay, which amounts to
parameterizing the effective interaction in terms of a few operators.
Any specific model would correspond to specific values for the
coupling constants in front of theses operators, and hence even a
rough measurement of these couplings could discriminate between
different NP models.

However, when determining the couplings in view of very sparse data
samples we are required to define proper observables, which we have
discussed in this paper. All observables are of the same nature as a
forward-backward asymmetry, and therefore can be measured by a simple
counting experiment.

On this basis we have performed a feasibility study on how precisely
one could assess the values of individual LFV couplings, based on only
a small number of total signal events. It turns out that for some
simplified cases (i.e.\ assuming short-distance coefficients to be real,
or particular chiral patterns) different NP scenarios could already be
distinguished with a quite small number of events. In the case that
LFV could be experimentally established, our procedure could be easily
extended by refining the binning for the energy phase space and by
including independent information on radiative $\tau \to \ell
\gamma$ decays.

\subsection*{Acknowledgments}

\noindent RB, TF, BOL and TM acknowledge support by the Deutsche
Forschungsgemeinschaft (DFG) within the Research Unit {\sc FOR~1873}
({\it Quark Flavour Physics and Effective Field Theories}).
{We are also grateful to Sven Faller who collaborated with us in the
early stages of this work.}
We further thank Danny van Dyk for helpful discussions.
The research of ST was supported by the ERC Advanced Grant EFT4LHC of the
European Research Council and the Cluster
of Excellence Precision Physics, Fundamental Interactions and Structure of
Matter (PRISMA-EXC 1098).
ST is deeply grateful for extensive discussions with Florian Bernlochner.

\newpage
\appendix

\noindent {\bf \Large Appendix}

\section{Operator matrix elements%
\label{apx:ME}}

Below we list contributions to the squared amplitude $|\mathcal
M^\uparrow|^2$. For legibility the symbol $\uparrow$ denoting that the
tau lepton in polarized spin-up in the $\vec s$ direction is
suppressed. We start with the contributions from squared leptonic
operators, which are
\begin{eqnarray}
  \label{eq:a1}
  |\mathcal M_{\rm lept}^{(LL)(LL)}|^2 &=& \left| \langle \mu^-(p_1) \mu^-(p_2)
    \mu^+(p_3) | H^{(LL)(LL)}_{\rm eff} | \tau^-(p_\tau) \rangle
  \right|^2 \nonumber \\
&=& \left|\xi_1\right|^2 32 (p_1 \cdot
    p_2) \left[ \mytauthree - m_\tau \mysthree \right] \;, \nonumber \\
|\mathcal M_{\rm lept}^{(RR)(RR)}|^2 &=& \left| \xi_2 \right|^2 32 (p_1 \cdot
p_2) \left[ \mytauthree + m_\tau \mysthree \right] \;, \nonumber \\
|\mathcal M_{\rm lept}^{(LL)(RR)}|^2 &=& \left| \xi_3 \right|^2 8 
\Big\{ (p_1 \cdot p_3) \left[ \mytautwo - m_\tau \mystwo \right]
+ (p_2 \cdot p_3) \left[ \mytauone - m_\tau \mysone \right] \Big.
\nonumber \\
&& \Big. \quad - m^2 \left[  \mytauthree  -  m_\tau \mysthree \right]
\Big\} \;, \nonumber \\
|\mathcal M_{\rm lept}^{(RR)(LL)}|^2 &=& \left| \xi_4 \right|^2 8 
\Big\{ (p_1 \cdot p_3) \left[ \mytautwo + m_\tau \mystwo \right]
+ (p_2 \cdot p_3) \left[ \mytauone + m_\tau \mysone \right] \Big.
\nonumber \\
&& \Big. \quad - m^2 \left[  \mytauthree  +  m_\tau \mysthree \right]
\Big\} \;.
\end{eqnarray}
Next, the interference terms from leptonic operators, which are
\begin{eqnarray}
  \label{eq:a2}
  \mathcal M_{\rm lept}^{(LL)(LL)} {\mathcal M_{\rm
      lept}^{(RR)(RR)}}^*  &=& \xi_1 \xi_2^* \,(-32) m_\tau m^3 \;,
  \nonumber \\
\mathcal M_{\rm lept}^{(LL)(LL)} {\mathcal M_{\rm lept}^{(LL)(RR)}}^*
&=& \xi_1 \xi_3^* \,8 m^2 \left[
  m_\tau \mysthree - \mytauthree + m_\tau^2 \right]\;,
  \nonumber \\
\mathcal M_{\rm lept}^{(LL)(LL)} {\mathcal M_{\rm lept}^{(RR)(LL)}}^*
&=& \xi_1 \xi_4^* \,8 m_\tau m
\left[ m_\tau \mysthree - \mytauthree + m^2 \right]\;,
  \nonumber \\
\mathcal M_{\rm lept}^{(RR)(RR)} {\mathcal M_{\rm lept}^{(LL)(RR)}}^*
&=& \xi_2 \xi_3^* \,8m_\tau m 
\left[  - m_\tau \mysthree - \mytauthree + m^2 \right]\;,
  \nonumber \\
\mathcal M_{\rm lept}^{(RR)(RR)} {\mathcal M_{\rm lept}^{(RR)(LL)}}^*
&=& \xi_2 \xi_4^* \,8 m^2 
\left[  - m_\tau \mysthree - \mytauthree + m_\tau^2 \right]\;,
  \nonumber \\
\mathcal M_{\rm lept}^{(LL)(RR)} {\mathcal M_{\rm lept}^{(RR)(LL)}}^*
&=& \xi_3 \xi_4^* \,8 m_\tau m
\left[ \myonetwo - 2 m^2 \right]\;,
\end{eqnarray}
as well as the complex conjugate expressions. The contributions from
radiative operators read
\begin{eqnarray}
  \label{eq:a3}
  |\mathcal M_{\rm rad}^{(LR)}|^2 &=& \left|\xi_5\right|^2  2 m_\tau^2 \Big\{ - 6
  m^2 \nonumber \\
&& + \frac{2m^2}{s_{13}^2} (m_\tau^2 - m^2) \left(m_\tau^2 -2 m_\tau
\mystwo - m^2\right) \nonumber \\
&& + \frac{2m^2}{s_{23}^2} (m_\tau^2 - m^2) \left(m_\tau^2 -2 m_\tau
\mysone - m^2\right) \nonumber \\
&& + \frac{2}{s_{13} s_{23}} (m_\tau^4 - 3 m_\tau^2 m^2+2 m^4)
\left( m^2 + m_\tau \mysone + m_\tau \mystwo \right) \nonumber \\
&& + \frac{1}{s_{13}} \Big[ \Big. 
m_\tau^4 + 3m^4 - 2(m_\tau^2 +3m^2) s_{23} + 2 s_{23}^2 \nonumber \\
&& \qquad \quad 
+ 4m_\tau (2m^2 - s_{23}) \mysone + 2m_\tau (3m^2 - 2m_\tau^2)
\mystwo \Big. \Big] \nonumber \\
&& + \frac{1}{s_{23}} \Big[ \Big. 
m_\tau^4 + 3m^4 - 2(m_\tau^2 +3m^2) s_{13} + 2 s_{13}^2 \nonumber \\
&& \qquad \quad + 4m_\tau (2m^2 - s_{13}) \mystwo + 2m_\tau (3m^2 - 2m_\tau^2)
\mysone \Big. \Big] \Big. \Big\} \;, \nonumber \\
|\mathcal M_{\rm rad}^{(RL)}|^2 &=& \left|\xi_6\right|^2  2 m_\tau^2 \Big\{ - 6
  m^2 \nonumber \\
&& + \frac{2m^2}{s_{13}^2} (m_\tau^2 - m^2) \left(m_\tau^2 + 2 m_\tau
\mystwo - m^2\right) \nonumber \\
&& + \frac{2m^2}{s_{23}^2} (m_\tau^2 - m^2) \left(m_\tau^2 + 2 m_\tau
\mysone - m^2\right) \nonumber \\
&& + \frac{2}{s_{13} s_{23}} (m_\tau^4 - 3 m_\tau^2 m^2+2 m^4)
\left( m^2 - m_\tau \mysone - m_\tau \mystwo \right) \nonumber \\
&& + \frac{1}{s_{13}} \Big[ \Big. 
m_\tau^4 + 3m^4 - 2(m_\tau^2 +3m^2) s_{23} + 2 s_{23}^2 \nonumber \\
&& \qquad \quad  - 4m_\tau (2m^2 - s_{23}) \mysone - 2m_\tau (3m^2 - 2m_\tau^2)
\mystwo \Big. \Big] \nonumber \\
&& + \frac{1}{s_{23}} \Big[ \Big. 
m_\tau^4 + 3m^4 - 2(m_\tau^2 +3m^2) s_{13} + 2 s_{13}^2 \nonumber \\
&& \qquad \quad  - 4m_\tau (2m^2 - s_{13}) \mystwo - 2m_\tau (3m^2 - 2m_\tau^2)
\mysone \Big. \Big] \Big. \Big\} \;, \nonumber \\
\mathcal M_{\rm rad}^{(LR)} {\mathcal M_{\rm rad}^{(RL)}}^* &=&
  \xi_5 \xi_6^*\, (-4 m_\tau^3 m) \left\{4 + 2 m^2
    \left(\frac{1}{s_{13}}+\frac{1}{s_{23}} \right) - 
    \frac{m^2( m_\tau^2 - m^2) }{s_{13}s_{23}} \right\} \;.
\end{eqnarray}
The mixed leptonic radiative contributions are the only ones that
contribute to $J_4$. We use the convention
\begin{equation}
  \label{eq:a4}
  \tr \left[ \gamma^\mu \gamma^\nu \gamma^\rho \gamma^\sigma \gamma_5
  \right] = - 4i \varepsilon^{\mu\nu\rho\sigma} \;.
\end{equation}
Since this trace is purely imaginary, we will pick up a dependence on
${\rm Im}(\xi_i \xi_j^*)$. We find
\begin{eqnarray}
  \label{eq:a5}
  \mathcal M_{\rm rad}^{(LR)} {\mathcal M_{\rm lept}^{(LL)(LL)}}^* &=&
  \xi_5 \xi_1^*\, 4 m_\tau  \Big\{ \Big.  
  2 m_\tau \left( m_\tau^2 - s_{13}-s_{23} \right) \nonumber \\
&& + m_\tau m^2 (m_\tau^2 - m^2) \left( \frac{1}{s_{13}} +
  \frac{1}{s_{23}} \right) \nonumber \\
&& +2 (m_\tau^2 + m^2 - s_{13}-s_{23} ) \left( \frac{1}{s_{13}} -
  \frac{1}{s_{23}} \right) i \myeps \nonumber \\
&& + (m_\tau^2 - m^2)(m_\tau^2 + m^2 - s_{23})
\frac{\mysone}{s_{13}} \nonumber \\
&& + (m_\tau^2 - m^2)(m_\tau^2 + m^2 - s_{13})
\frac{\mystwo}{s_{23}} \nonumber \\
&& + \left[ (m_\tau^2 - s_{13})^2 -2m^2 s_{13}+m^4 \right]
\frac{\mysone}{s_{23}} \nonumber \\
&& + \left[ (m_\tau^2 - s_{23})^2 -2m^2 s_{23}+m^4 \right]
\frac{\mystwo}{s_{13}} \nonumber \\
&& - \left( m_\tau^2 +s_{23}-m^2 \right) \mysone
- \left( m_\tau^2 +s_{13}-m^2 \right) \mystwo \Big. \Big\}\;.
\end{eqnarray}
Next,
\begin{eqnarray}
\label{eq:a6}
\mathcal M_{\rm rad}^{(LR)} {\mathcal M_{\rm lept}^{(RR)(RR)}}^* &=&
  \xi_5 \xi_2^* \,4 m_\tau  m \Big\{ \Big.   2 \left( m^2 - s_{13}
    - s_{23} \right)   + 4m_\tau \left[ \mysone+\mystwo \right]
  \nonumber \\
&& + \frac{m^2}{s_{13}} \left[ m^2 - m_\tau^2 + 2m_\tau\mystwo \right]
+ \frac{m^2}{s_{23}} \left[ m^2 - m_\tau^2 + 2m_\tau\mysone \right]
\Big. \Big\}\;.
\end{eqnarray}
Next,
\begin{eqnarray}
  \label{eq:a7}
  \mathcal M_{\rm rad}^{(LR)} {\mathcal M_{\rm lept}^{(LL)(RR)}}^* &=&
  \xi_5 \xi_3^* 2 m_\tau  \Big\{ \Big. m_\tau \left(
    s_{13}+s_{23}-6m^2 \right) \nonumber \\
&& + m_\tau m^2 (m_\tau^2 - m^2) \left( \frac{1}{s_{13}}+
  \frac{1}{s_{23}} \right) \nonumber \\
&& - 2 \left( \frac{s_{23}}{s_{13}} - \frac{s_{13}}{s_{23}} \right)
i\myeps \nonumber \\
&& - (m_\tau^2-m^2) \frac{s_{23}}{s_{13}} \mysone
- (m_\tau^2-m^2) \frac{s_{13}}{s_{23}} \mystwo \nonumber \\
&& - s_{23} \mysone - s_{13} \mystwo \nonumber \\
&& - \left[ (m_\tau^2 - s_{13}) s_{13} + m^2 (2m_\tau^2 + s_{13})
\right] \frac{\mysone}{s_{23}} \nonumber \\
&& - \left[ (m_\tau^2 - s_{23}) s_{23} + m^2 (2m_\tau^2 + s_{23})
\right] \frac{\mystwo}{s_{13}} \Big. \Big\}.
\end{eqnarray}
Next,
\begin{eqnarray}
  \label{eq:a8}
  \mathcal M_{\rm rad}^{(LR)} {\mathcal M_{\rm lept}^{(RR)(LL)}}^* &=&
  \xi_5 \xi_4^* 4 m_\tau m  \Big\{ \Big. m^2 - s_{13} - s_{23}
  \nonumber \\
&& - \frac{m^2}{2} (m_\tau^2 - m^2) 
\left( \frac{1}{s_{13}} +  \frac{1}{s_{23}} \right) \nonumber \\
&& - 2m_\tau \left( \frac{1}{s_{13}} -  \frac{1}{s_{23}} \right) i
\myeps \nonumber \\
&& - m_\tau (m_\tau^2 - m^2) \frac{\mysone}{s_{13}}
- m_\tau (m_\tau^2 - m^2) \frac{\mystwo}{s_{23}} \nonumber \\
&& - m_\tau (m_\tau^2 - s_{13}) \frac{\mysone}{s_{23}}
- m_\tau (m_\tau^2 - s_{23}) \frac{\mystwo}{s_{13}} \nonumber \\
&& + m_\tau \left[ \mysone+\mystwo \right] \Big. \Big\} \;.
\end{eqnarray}
Now we mix the second radiative operator into the leptonic ones. The
expressions are the same as we found before, except for the signs in
front of the spin-vector products $\mysone$ and $\mystwo$. 
\begin{eqnarray}
  \label{eq:a9}
    \mathcal M_{\rm rad}^{(RL)} {\mathcal M_{\rm lept}^{(LL)(LL)}}^* &=&
  \xi_6 \xi_1^* 4 m_\tau  m \Big\{ \Big.   2 \left( m^2 - s_{13}
    - s_{23} \right)   - 4m_\tau \left[ \mysone+\mystwo \right]
  \nonumber \\
&& + \frac{m^2}{s_{13}} \left[ m^2 - m_\tau^2 - 2m_\tau\mystwo \right]
+ \frac{m^2}{s_{23}} \left[ m^2 - m_\tau^2 - 2m_\tau\mysone \right]
\Big. \Big\}\;.
\end{eqnarray}
Next,
\begin{eqnarray}
  \label{eq:a10}
    \mathcal M_{\rm rad}^{(RL)} {\mathcal M_{\rm lept}^{(RR)(RR)}}^* &=&
  \xi_6 \xi_2^* 4 m_\tau  \Big\{ \Big.  
  2 m_\tau \left( m_\tau^2 - s_{13}-s_{23} \right) \nonumber \\
&& + m_\tau m^2 (m_\tau^2 - m^2) \left( \frac{1}{s_{13}} +
  \frac{1}{s_{23}} \right) \nonumber \\
&& +2 (m_\tau^2 + m^2 - s_{13}-s_{23} ) \left( \frac{1}{s_{13}} -
  \frac{1}{s_{23}} \right) i \myeps \nonumber \\
&& - (m_\tau^2 - m^2)(m_\tau^2 + m^2 - s_{23})
\frac{\mysone}{s_{13}} \nonumber \\
&& - (m_\tau^2 - m^2)(m_\tau^2 + m^2 - s_{13})
\frac{\mystwo}{s_{23}} \nonumber \\
&& - \left[ (m_\tau^2 - s_{13})^2 -2m^2 s_{13}+m^4 \right]
\frac{\mysone}{s_{23}} \nonumber \\
&& - \left[ (m_\tau^2 - s_{23})^2 -2m^2 s_{23}+m^4 \right]
\frac{\mystwo}{s_{13}} \nonumber \\
&& + \left( m_\tau^2 +s_{23}-m^2 \right) \mysone
+ \left( m_\tau^2 +s_{13}-m^2 \right) \mystwo \Big. \Big\}\;.
\end{eqnarray}
Next,
\begin{eqnarray}
  \label{eq:a11}
    \mathcal M_{\rm rad}^{(RL)} {\mathcal M_{\rm lept}^{(LL)(RR)}}^* &=&
  \xi_6 \xi_3^* 4 m_\tau m  \Big\{ \Big. m^2 - s_{13} - s_{23}
  \nonumber \\
&& - \frac{m^2}{2} (m_\tau^2 - m^2) 
\left( \frac{1}{s_{13}} +  \frac{1}{s_{23}} \right) \nonumber \\
&& - 2m_\tau \left( \frac{1}{s_{13}} -  \frac{1}{s_{23}} \right) i
\myeps \nonumber \\
&& + m_\tau (m_\tau^2 - m^2) \frac{\mysone}{s_{13}}
+ m_\tau (m_\tau^2 - m^2) \frac{\mystwo}{s_{23}} \nonumber \\
&& + m_\tau (m_\tau^2 - s_{13}) \frac{\mysone}{s_{23}}
+ m_\tau (m_\tau^2 - s_{23}) \frac{\mystwo}{s_{13}} \nonumber \\
&& - m_\tau \left[ \mysone+\mystwo \right] \Big. \Big\}
\end{eqnarray}
And finally
\begin{eqnarray}
  \label{eq:a12}
    \mathcal M_{\rm rad}^{(RL)} {\mathcal M_{\rm lept}^{(RR)(LL)}}^* &=&
  \xi_6 \xi_4^* 2 m_\tau  \Big\{ \Big. m_\tau \left(
    s_{13}+s_{23}-6m^2 \right) \nonumber \\
&& + m_\tau m^2 (m_\tau^2 - m^2) \left( \frac{1}{s_{13}}+
  \frac{1}{s_{23}} \right) \nonumber \\
&& - 2 \left( \frac{s_{23}}{s_{13}} - \frac{s_{13}}{s_{23}} \right)
i\myeps \nonumber \\
&& + (m_\tau^2-m^2) \frac{s_{23}}{s_{13}} \mysone
+ (m_\tau^2-m^2) \frac{s_{13}}{s_{23}} \mystwo \nonumber \\
&& + s_{23} \mysone + s_{13} \mystwo \nonumber \\
&& + \left[ (m_\tau^2 - s_{13}) s_{13} + m^2 (2m_\tau^2 + s_{13})
\right] \frac{\mysone}{s_{23}} \nonumber \\
&& + \left[ (m_\tau^2 - s_{23}) s_{23} + m^2 (2m_\tau^2 + s_{23})
\right] \frac{\mystwo}{s_{13}} \Big. \Big\}.
\end{eqnarray}

\newpage 

\section{Analytic Calculation of the Matrices $\bm{A}_i^{\rm (R)}$}

\label{apx:MR}

In the following, we show two examples how to obtain analytic results
for the matrices $A_i^{(R)}$ as an expansion in terms of the small
parameter $\ee = m/m_\tau$, keeping terms up to  $\mathcal O(\ee^2)$.

\paragraph{Integrating a constant over the full phase space. } We
  integrate the constant $1/m_\tau^2$ over the energies $E_2$ and
  $E_3$. At leading power in $\ee$ there is no problem and the result
  is $1/8$. However, at subleading power divergences appear at the
  border of the phase space. We regulate these divergencies by
  manually introducing 
  \begin{equation}
    \label{eq:29}
      \frac{1}{m_\tau^2} \longrightarrow \left( \frac{E_2}{\mu_1}
  \right)^{\eta_1} \left( \frac{s_{13}}{\mu_2^2} \right)^{\eta_2} \frac{1}{m_\tau^2} \;,
  \end{equation}
  and taking the simultaneous limit $\eta_{1,2}\to 0$ in the end. The
  first integration -- over $E_3$, say -- is straight forward. We then
  distinguish three regions.
  \begin{itemize}
  \item Treating $E_2$ as $\mathcal O(m_\tau)$ yields
    \begin{equation}
      \label{eq:30}
        I_{\rm hard} = \frac18 \underbrace{-\frac{m^2}{2m_\tau^2
      \eta_1}-\frac{m^2}{2m_\tau^2 \eta_2}}_{\rm singular} +
  \frac{m^2}{m_\tau^2} \left[ \frac34 - \left( \frac12+
      \frac{\eta_1}{2\eta_2} \right) \ln \frac{m_\tau}{2\mu_1} 
  - \left(1+ \frac{\eta_2}{\eta_1} \right) \ln \frac{m_\tau}{\mu_2}
\right] \;.
    \end{equation}
  \item When treating $E_2 \sim m$ one needs to
    integrate $E_2 \in [m,\infty]$. This gives
    \begin{equation}
      \label{eq:31}
        I_{\rm soft\;E_2} = \underbrace{\frac{m^2}{2m_\tau^2
      \eta_1}}_{\rm singular} + \frac{m^2}{m_\tau^2} \left[  -\frac14
    + \frac12 \ln \frac{m}{2\mu_1} + \frac{\eta_2}{\eta_1} \ln
    \frac{m_\tau}{\mu_2} \right]\;.
    \end{equation}
  \item $E_2$ is near its maximum value is akin to treating $s_{13}
    \sim m^2$ and integrating $s_{13} \in [4m^2,\infty]$. We find
    \begin{equation}
      \label{eq:32}
      I_{\rm soft\;s_{13}} = \underbrace{\frac{m^2}{2m_\tau^2
      \eta_2}}_{\rm singular} + \frac{m^2}{m_\tau^2} \left[  -\frac12
    + \ln \frac{m}{\mu_2} + \frac{\eta_1}{2\eta_2} \ln
    \frac{m_\tau}{2\mu_1} \right]\;.
    \end{equation}
  \end{itemize}
  The singular terms cancel in the sum of these contributions, and the
  dependence on the auxiliary scales $\mu_i$ drops out as well,
  resulting in
  \begin{equation}
    \label{eq:33}
     I_{\rm hard} +  I_{\rm soft\;E_2} +  I_{\rm
    soft\;s_{13}} = \frac18 + \frac{3m^2}{2m_\tau^2} \ln
  \frac{m}{m_\tau} .
  \end{equation}
\paragraph{Integrating singular pieces from the radiative
    contributions. } Much more challenging contributions arise from the
  radiative operators, where the intermediate photon propagator leads
  to integrands proportional to inverse powers of the invariants
  $s_{ij}$.  The most singular expression in (\ref{eq:a3}), for
  example, is 
  \begin{equation}
    \label{eq:34}
    \frac{1}{s_{13}s_{23}} = \frac{1}{(s_{13}+s_{23})s_{13}} +
\frac{1}{(s_{13}+s_{23})s_{23}} \;.
  \end{equation}
  For definiteness we are therefore focussing here on the integral
  \begin{equation}
    \label{eq:35}
      S(m/m_\tau) = \int\limits_{\rm phase\;space} dE_2\,dE_3
  \frac{m_\tau^2}{(s_{13}+s_{23})s_{13}}
  = \int\limits_{\rm phase\;space} dE_2\,dE_3
  \frac{m_\tau^2}{(2 m_\tau E_3 + 2m^2)s_{13}}\;,
  \end{equation}
  which has a soft $E_3$ and a collinear $s_{13}$ singular
  structure. We may identify four different regions to solve this
  integral.
  \begin{enumerate}
  \item The soft $E_2$ region, where $E_2 \sim m$ and $b_2$ can be
    resolved. This region only becomes relevant for power corrections.
  \item The collinear $s_{13} \sim m^2$ region, where $E_2$ is nearly at the endpoint,
    namely of the order $\mathcal O(m^2/m_\tau)$ away from its
    maximum. Here $d_2$ can be resolved.
  \item The soft $E_3$ region, where $E_2$ is of order $\mathcal O(m)$
    away from its maximum.
  \item The bulk.
  \end{enumerate}
  We aim to distinguish these regions in a single, one-dimensional
  integral over a variable that differentiates between the four
  regions in terms of its power counting. To this end we perform the
  first integration over $E_3$ without approximation and are left with
  a single integral over $E_2$. Let us now define $\tilde E_2$ as the
  energy variable that is smallest for maximal $E_2$, i.e.~$E_2
  = \frac{m_\tau^2-3m^2}{2m_\tau} - \tilde E_2$. Finally we define the
  variable $\rho = \tilde E_2 / E_2$, which will accomplish our goal:
  After substituting the $E_2 =  \frac{m_\tau^2-3m^2}{2m_\tau}
  \frac{1}{1+\rho}$ integration variable for $\rho \in
  [0,\frac{m_\tau}{2m} - 1 - \frac{3m}{2m_\tau}]$ we find the
  following in the above  four regions and name them
  \begin{enumerate}
  \item $\rho \sim m_\tau / m$, label ``uhard'',
  \item $\rho \sim m^2 / m_\tau^2$, label ``usoft'',
  \item $\rho \sim m/m_\tau$, label ``soft'',
  \item $\rho \sim 1$, label ``hard''.
  \end{enumerate}
  Finally we need to choose a regulator common in all regions. While
  $\rho^\eta \stackrel{\eta \to 0}{\longrightarrow} 1$ is quite
  suitable we find that the regulator
  \begin{equation}
    \label{eq:36}
    \left(
    \frac{\sqrt{s_{13}}+\sqrt{s_{13}-4m^2}}{\sqrt{s_{13}}-\sqrt{s_{13}-4m^2}}
    \frac{m^2}{\mu_1^2} \right)^\eta
  \left( \frac{E_2}{\mu_2} \right)^{2\eta} \stackrel{\eta \to
    0}{\longrightarrow} 1 
  \end{equation}
  is more advantageous as it simplifies the practical calculation. We
  find
  \begin{itemize}
  \item {\bf Usoft region:} Expanding the integrand and
integrating $0\le \rho < \infty$ we find
\begin{equation}
  \label{eq:37}
  S_{us} = \frac{1}{4\eta^2} + \frac{1}{2\eta} \left[ \ln \frac{m_\tau
    m}{2\mu_1 \mu_2} - \frac{2m^2}{m_\tau^2} \right] + \hbox{finite.}
\end{equation}
\item {\bf Soft region:} Expanding the integrand and
integrating $0\le \rho < \infty$ we find
\begin{equation}
  \label{eq:38}
  S_s = - \frac{1}{2\eta^2} + \frac{1}{2\eta} \left[ \ln \frac{4\mu_1^2
      \mu_2^2}{m_\tau^3 m} + \frac{m^2}{2m_\tau^2} \right] + \hbox{finite.}
\end{equation}
\item {\bf Hard region:} Expanding the integrand and
integrating $0\le \rho < \infty$ we find
\begin{equation}
  \label{eq:39}
S_h = \frac{1}{4\eta^2} + \frac{1}{2\eta} \left[ \ln
    \frac{m_\tau^2}{2\mu_1 \mu_2} + \frac{m^2}{m_\tau^2} \right] + \hbox{finite.}
\end{equation}
\item {\bf Uhard region:} Expanding the integrand and
integrating $0 \le \rho \le \frac{m_\tau}{2m} - 1 -
\frac{3m}{2m_\tau}$ we find
\begin{equation}
  \label{eq:40}
S_{uh} = \frac{1}{4\eta} \frac{m^2}{m_\tau^2}  + \hbox{finite.}
\end{equation}
  \end{itemize}
Again all the double and single divergences cancel in the sum of the
regions and we obtain a leading double logarithm free of any
regulators, to wit
\begin{equation}
  \label{eq:41}
  S(m/m_\tau) = \left[ \frac14 \ln^2 \frac{m}{m_\tau} - \frac{\pi^2}{16}\right] +
  \frac{m^2}{m_\tau^2} \left[ \frac98 - \frac54 \ln \frac{m}{m_\tau} \right]\;.
\end{equation}

Armed with the techniques discussed above we are able to state some
analytic results. We find in terms of $\ee = m/m_\tau$
\begin{equation}
  \label{eq:42}
\bm{A}_1^{\rm \{full\}} = 
  \left( \begin{array}{cccccc}
\bm{1}-24\ee^2 & 0 & 4\ee^2 & -2\ee & \bm{2}-48\ee^2 & -4\ee \\
0 & \bm{1}-24\ee^2 & -2\ee & 4\ee^2 &  -4\ee & \bm{2}-48\ee^2 \\
4\ee^2 &-2\ee & \bm{\frac12} - 14\ee^2 & \ee & \bm{1}-24\ee^2 & -2\ee\\
-2\ee &4\ee^2& \ee & \bm{\frac12} - 14\ee^2 & -2\ee & \bm{1}-24\ee^2 \\
\bm{2}-48\ee^2 &-4\ee& \bm{1}-24\ee^2 & -2\ee & \bm{a^{\rm \{full\}}_{55}} & -12\ee \\
-4\ee & \bm{2}-48\ee^2 & -2\ee & \bm{1}-24\ee^2 & -12\ee & \bm{a^{\rm \{full\}}_{66}}
\end{array} \right) + \mathcal O(\ee^3) \;,
\end{equation}
where
\begin{equation}
  \label{eq:43}
  a^{\rm \{full\}}_{55} = a^{\rm \{full\}}_{66} = - (8 \ln \ee + 11) + \left[ 12 \ln^2\ee + 36
    \ln\ee + 50-3\pi^2 \right] \ee^2 \;.
\end{equation}

\newpage
\section{Numerical Results for the Matrices $\bm{A}_i^{\rm (R)}$}
\label{apx:2}

For muons in the final state we use $m = 105.658$~MeV and $m_\tau =
1776.82$~MeV, which corresponds to $\ee = 0.0595$. 
\begin{eqnarray}
  \label{eq:a13}
  \bm{A}_1^{\rm \{full\}} &\simeq& \left(
\begin{array}{cccccc}
  0.919 & -0.004 & 0.012 & -0.104 & 1.838 & -0.217 \\
  -0.004 & 0.919 & -0.104 & 0.012 & -0.217 & 1.838 \\
  0.012 & -0.104 & 0.453 & 0.050 & 0.919 & -0.109 \\
  -0.104 & 0.012 & 0.050 & 0.453 & -0.109 & 0.919 \\
  1.838 & -0.217 & 0.919 & -0.109 & 11.635 & -0.628 \\
  -0.217 & 1.838 & -0.109 & 0.919 & -0.628 & 11.635 
\end{array} \right) \nonumber \\
  \bm{A}_2^{\rm \{full\}} &\simeq & \left(
\begin{array}{cccccc}
  0.889 & 0           & -0.006 & -0.103 & 1.461 & -0.210 \\
          0 & -0.889 & 0.103 & 0.006 & 0.210 & -1.461 \\
  -0.006 & 0.103 & -0.159 & 0    & -0.848 & 0.185 \\
  -0.103 & 0.006 & 0      & 0.159 & -0.185 & 0.848 \\
  1.461 & 0.210 & -0.848 & -0.185 & -4.344 & 0 \\
  -0.210 & -1.461 & 0.185 & 0.848 & 0 & 4.344 
\end{array} \right) \nonumber \\
  \bm{A}_3^{\rm \{full\}} &=& 
  \bm{A}_4^{\rm \{full\}} =0
\end{eqnarray}

For the part of the phase space where $\cos\phi_{23}>0$ we find
\begin{eqnarray}
  \label{eq:a14}
  \bm{A}_1^{\{\cos\phi_{23}>0 \}} &\simeq& \left(
\begin{array}{cccccc}
    0.245 & -0.001 & 0.003 & -0.019 & 0.572 & -0.041 \\
    -0.001 & 0.245 & -0.019 & 0.003 & -0.041 & 0.572 \\
    0.003 & -0.019 & 0.081 & 0.015 & 0.175 & -0.020 \\
    -0.019 & 0.003 & 0.015 & 0.081 & -0.020 & 0.176 \\
    0.572 & -0.041 & 0.175 & -0.020 & 4.320 & -0.139 \\
    -0.041 & 0.572 & -0.020 & 0.175 & -0.139 & 4.320
\end{array} \right)\;, \nonumber \\
  \bm{A}_2^{\{\cos\phi_{23}>0 \}} &\simeq& \left(
\begin{array}{cccccc}
  0.235 & 0 & -0.001 & -0.018 & 0.417 & -0.038 \\
  0 & -0.235 & 0.018 & 0.001 & 0.038 & -0.417 \\
  -0.001 & 0.018 & 0.009 & 0 & -0.145 & 0.041 \\
  -0.018 & 0.001 & 0 & -0.009 & -0.041 & 0.145 \\
  0.417 & 0.038 & -0.145 & -0.041 & -1.831 & 0 \\
  -0.038 & -0.417 & 0.041 & 0.145 & 0 & 1.831
\end{array} \right)\;, 
\end{eqnarray}
\begin{eqnarray}
  \bm{A}_3^{\{\cos\phi_{23}>0 \}} &\simeq& \left(
\begin{array}{cccccc}
  0 & 0 & 0 & 0 & -0.276 & -0.0005 \\
  0 & 0 & 0 & 0 & 0.0005 & 0.276 \\
  0 & 0 & 0.037 & 0 & 0.162 & -0.036 \\
  0 & 0 & 0 & -0.037 & 0.036 & -0.162 \\
  -0.276 & 0.0005 & 0.162 & 0.036 & -1.489 & 0 \\
  -0.0005 & 0.276 & -0.036 & -0.162 & 0 & 1.489
\end{array} \right)\;, \nonumber \\
  \bm{A}_4^{\{\cos\phi_{23}>0 \}} &\simeq& \left(
\begin{array}{cccccc}
  0 & 0 & 0 & 0 & 0.253 i & 0 \\
  0 & 0 & 0 & 0 & 0 & 0.253 i \\
  0 & 0 & 0 & 0 & -0.162 i & -0.034 i \\
  0 & 0 & 0 & 0 & -0.034 i & -0.162 i \\
  -0.253 i & 0 & 0.162 i & 0.034 i & 0 & 0 \\
  0 & -0.253 i & 0.034 i & 0.162 i & 0 & 0
\end{array} \right)\;,
\end{eqnarray}
and for R$=\{\cos\phi_{12}>0\}$ we find
\begin{eqnarray}
  \label{eq:a15}
  \bm{A}_1^{\{\cos\phi_{12}>0 \}} &\simeq& \left(
\begin{array}{cccccc}
0.083 & -0.001 & 0.002 & -0.032 & 0.100 & -0.065 \\
-0.001 & 0.083 & -0.032 & 0.002 & -0.065 & 0.100 \\
0.002 & -0.032 & 0.121 & 0.002 & 0.272 & -0.033 \\
-0.032 & 0.002 & 0.002 & 0.121 & -0.033 & 0.272 \\
0.100 & -0.065 & 0.272 & -0.033 & 1.001 & -0.141 \\
-0.065 & 0.100 & -0.033 & 0.272 & -0.141 & 1.001
\end{array} \right)\;, \nonumber \\
  \bm{A}_2^{\{\cos\phi_{12}>0 \}} &\simeq& \left(
\begin{array}{cccccc}
0.082 & 0 & -0.002 & -0.032 & 0.084 & -0.064 \\
0 & -0.082 & 0.032 & 0.002 & 0.064 & -0.084 \\
-0.002 & 0.032 & -0.098 & 0 & -0.269 & 0.037 \\
-0.032 & 0.002 & 0 & 0.098 & -0.037 & 0.269 \\
0.084 & 0.064 & -0.269 & -0.037 & -0.793 & 0 \\
-0.064 & -0.084 & 0.037 & 0.269 & 0 & 0.793
\end{array} \right)\;, \nonumber \\
\bm{A}_3^{\{\cos\phi_{12}>0 \}} &=& \bm{A}_4^{\{\cos\phi_{12}>0 \}} = 0\;.
\end{eqnarray}

\newpage
Now we repeat the calculation for electrons in the final state. Here
$m=510.9989$ KeV. This means that $\ee = 0.0002876$ is much smaller
and we can observe the structure of the matrices much easier. We find
\begin{eqnarray}
  \label{eq:a16}
  \bm{A}_1^{\rm \{full\}}&=& \left(
\begin{array}{cccccc}
1.00 & 0 & 0 & 0 & 2.00 & 0 \\
0 & 1.00 & 0 & 0 & 0 & 2.00 \\
0 & 0 & 0.50 & 0 & 1.00 & 0 \\
0 & 0 & 0 & 0.50 & 0 & 1.00 \\
2.00 & 0 & 1.00 & 0 & 54.23 & 0 \\
0 & 2 & 0 & 1.00 & 0 & 54.23 
\end{array} \right) + \mathcal O(10^{-3})\;, \nonumber \\
  \bm{A}_2^{\rm \{full\}}&=& \left(
\begin{array}{cccccc}
1.00 & 0 & 0 & 0 & 2.00 & 0  \\
0 & -1.00 & 0 & 0 & 0 & -2.00 \\
0 & 0 & -0.17 & 0 & -1.00 & 0 \\
0 & 0 & 0 & 0.17 & 0 & 1.00 \\
2.00 & 0 & -1.00 & 0 & -42.26 & 0 \\
0 & -2.00 & 0 & 1.00 & 0 & 42.26 
\end{array} \right)  + \mathcal O(10^{-3})\;, \nonumber \\
  \bm{A}_3^{\rm \{full\}} &=& \bm{A}_4^{\rm \{full\}} = 0\;,
\end{eqnarray}
\begin{eqnarray}
  \label{eq:a17}
  \bm{A}_1^{\{\cos\phi_{23}>0 \}} &=& \left(
\begin{array}{cccccc}
0.27 & 0 & 0 & 0 & 0.64 & 0 \\
0 & 0.27 & 0 & 0 & 0 & 0.64 \\
0 & 0 & 0.09 & 0 & 0.18 & 0 \\
0 & 0 & 0 & 0.09 & 0 & 0.18 \\
0.64 & 0 & 0.18 & 0 & 25.22 & 0 \\
0 & 0.64 & 0 & 0.18 & 0 & 25.22 \\
\end{array} \right)  + \mathcal O(10^{-3})\;, \nonumber \\
  \bm{A}_2^{\{\cos\phi_{23}>0 \}} &=& \left(
\begin{array}{cccccc}
0.27 & 0 & 0 & 0 & 0.64 & 0 \\
0 & -0.27 & 0 & 0 & 0 & -0.64 \\
0 & 0 & 0.01 & 0 & -0.18 & 0 \\
0 & 0 & 0 & -0.01 & 0 & 0.18 \\
0.64 & 0 & -0.18 & 0 & -20.60 & 0 \\
0 & -0.64 & 0 & 0.18 & 0 & 20.60 \\
\end{array} \right)  + \mathcal O(10^{-3})\;, 
\end{eqnarray}
\begin{eqnarray}
  \bm{A}_3^{\{\cos\phi_{23}>0 \}} &=& \left(
\begin{array}{cccccc}
0 & 0 & 0 & 0 & -0.63 & 0 \\
0 & 0 & 0 & 0 & 0 & 0.63 \\
0 & 0 & 0.04 & 0 & 0.36 & 0 \\
0 & 0 & 0 & -0.04 & 0 & -0.36 \\
-0.63 & 0 & 0.36 & 0 & -4.57 & 0 \\
0 & 0.63 & 0 & -0.36 & 0 & 4.57 \\
\end{array} \right)  + \mathcal O(10^{-3})\;, \nonumber \\
  \bm{A}_4^{\{\cos\phi_{23}>0 \}} &=& \left(
\begin{array}{cccccc}
0 & 0 & 0 & 0 & 0.63i & 0 \\
0 & 0 & 0 & 0 & 0 & 0.63i \\
0 & 0 & 0 & 0 & -0.36i & 0 \\
0 & 0 & 0 & 0 & 0 & -0.36i \\
-0.63i & 0 & 0.36i & 0 & 0 & 0 \\
0 & -0.63i & 0 & 0.36i & 0 & 0 \\
\end{array} \right)  + \mathcal O(10^{-3})\;, 
\end{eqnarray}
\begin{eqnarray}
  \label{eq:a18}
  \bm{A}_1^{\{\cos\phi_{12}>0 \}} &=& \left(
\begin{array}{cccccc}
0.08 & 0 & 0 & 0 & 0.09 & 0 \\
0 & 0.08 & 0 & 0 & 0 & 0.09 \\
0 & 0 & 0.14 & 0 & 0.32 & 0 \\
0 & 0 & 0 & 0.14 & 0 & 0.32 \\
0.09 & 0 & 0.32 & 0 & 1.52 & 0 \\
0 & 0.09 & 0 & 0.32 & 0 & 1.52 \\
\end{array} \right)  + \mathcal O(10^{-3})\;, \\
  \bm{A}_2^{\{\cos\phi_{12}>0 \}} &=& \left(
\begin{array}{cccccc}
0.08 & 0 & 0 & 0 & 0.09 & 0 \\
0 & -0.08 & 0 & 0 & 0 & -0.09 \\
0 & 0 & -0.11 & 0 & -0.32 & 0 \\
0 & 0 & 0 & 0.11 & 0 & 0.32 \\
0.09 & 0 & -0.32 & 0 & -1.32 & 0 \\
0 & -0.09 & 0 & 0.32 & 0 & 1.32 \\
\end{array} \right)  + \mathcal O(10^{-3})\;, \\
\bm{A}_3^{\{\cos\phi_{12}>0 \}} &=& \bm{A}_4^{\{\cos\phi_{12}>0 \}} = 0\;.
\end{eqnarray}

\end{document}